\documentclass[aps,prb,superscriptaddress,longbibliography, twocolumn, floatfix]{revtex4-2}
\usepackage{bm}
\usepackage{lmodern}

\usepackage[utf8]{inputenc}
\usepackage{amsmath}
\usepackage{nicematrix}
\NiceMatrixOptions{
code-for-first-row = \color{blue} ,
code-for-last-row = \color{blue} ,
code-for-first-col = \color{blue} ,
code-for-last-col = \color{blue}
}
\usepackage[utf8]{inputenc} 

\usepackage{mathtools}
\usepackage{siunitx}
\usepackage{amssymb}
\usepackage{graphicx}
\usepackage{epstopdf}
\usepackage{xcolor}
\usepackage{enumerate}
\usepackage{ulem}
\usepackage[english]{babel}
\usepackage{braket}
\usepackage{natbib}
\usepackage{svg}
\usepackage{float}
\usepackage{stmaryrd}
\usepackage[T1]{fontenc}

\usepackage{hyperref}

\usepackage{cleveref}
\usepackage{lipsum}

\usepackage[title]{appendix}

\begin{document}

\title{A Three-Dimensional Array of Quantum Dots}

\author{Hanifa Tidjani}\altaffiliation{These authors contributed equally}
\affiliation{QuTech and Kavli Institute of Nanoscience, Delft University of Technology, PO Box 5046, 2600 GA Delft, The Netherlands}
\author{Dario Denora}\altaffiliation{These authors contributed equally}
\affiliation{QuTech and Kavli Institute of Nanoscience, Delft University of Technology, PO Box 5046, 2600 GA Delft, The Netherlands}
\author{Michael Chan}
\affiliation{QuTech and Kavli Institute of Nanoscience, Delft University of Technology, PO Box 5046, 2600 GA Delft, The Netherlands}

\author{Jann Hinnerk Ungerer}
\affiliation{QuTech and Kavli Institute of Nanoscience, Delft University of Technology, PO Box 5046, 2600 GA Delft, The Netherlands}
\affiliation{Department of Physics, Harvard University, Cambridge, MA, USA}

\author{Barnaby van Straaten}
\affiliation{QuTech and Kavli Institute of Nanoscience, Delft University of Technology, PO Box 5046, 2600 GA Delft, The Netherlands}

\author{Stefan D. Oosterhout}
\affiliation{Netherlands Organisation for Applied Scientific Research (TNO), Delft, The Netherlands}

\author{Lucas Stehouwer}
\affiliation{QuTech and Kavli Institute of Nanoscience, Delft University of Technology, PO Box 5046, 2600 GA Delft, The Netherlands}

\author{Giordano Scappucci}
\affiliation{QuTech and Kavli Institute of Nanoscience, Delft University of Technology, PO Box 5046, 2600 GA Delft, The Netherlands}
\author{Menno Veldhorst}
\affiliation{QuTech and Kavli Institute of Nanoscience, Delft University of Technology, PO Box 5046, 2600 GA Delft, The Netherlands}
\date{\today}

\begin{abstract}

Quantum dots can confine single electrons or holes to define spin qubits that can be operated with high fidelity. Experimental work has progressed from linear to two-dimensional arrays of quantum dots, enabling qubit interactions that are essential for quantum simulation and computation. Here, we explore architectures beyond planar geometries by constructing quantum dot arrays in three dimensions. We realize an eight-quantum dot system in a silicon-germanium heterostructure comprising two stacked germanium quantum wells, where quantum dots are positioned at the vertices of a cuboid. Using electrostatic gate control, we load a single hole into any of the eight quantum dots. To demonstrate the potential of multilayer quantum dot systems, we show coherent spin control and hopping-induced spin rotations by shuttling between the quantum wells. The ability to extend quantum dot arrays in three dimensions provides opportunities for novel quantum hardware and high-connectivity quantum circuits.

\end{abstract}

\maketitle
\section*{Introduction}
Interacting quantum systems can give rise to a wide range of phenomena. Predicting their behavior remains a formidable challenge and is considered one of the first practical applications of quantum computers~\cite{Feynman1982}. Quantum simulation may be particularly powerful when there exists a direct mapping. For example, a two-dimensional quantum simulator with nearest-neighbor interactions may be used to explore non-trivial phase transitions in the Fermi-Hubbard model~\cite{fauseweh2024quantum}. However, real systems extend in three dimensions, and thus even higher connectivity may be needed for simulations of nature~\cite{shao2024antiferromagnetic}.

Quantum connectivity is also highly relevant for quantum computation and has a large impact on the performance and efficiency of quantum error correction schemes. Two-dimensional arrays with nearest-neighbor interactions may support surface codes with error thresholds as high as 1\%~\cite{raussendorf2007fault, fowler2012surface}. Even higher connectivity may drastically lower the physical qubit overhead to construct logical qubits~\cite{Terhal2015QuantumMemories}.


Atomic qubits such as trapped ions and neutral atoms can have high connectivity, due to the use of collective vibrational modes, by shuttling, or by arranging qubits in three dimensions~\cite{cirac1995quantum, Ashkin1997, Barredo2018}. In the solid state, achieving high connectivity is more challenging. For example, spin qubits in semiconductor quantum dots have only recently been scaled into small systems with two-dimensional connectivity~\cite{Hendrickx2021, Wang2024, John2024, unseld2024baseband}. 


A first route to further increase the connectivity between spin qubits could be to exploit quantum links based on shuttling~\cite{taylor2005fault} or superconducting resonators~\cite{trif2008spin}. A complementary route would be the construction of quantum dot arrays with a higher connectivity. Advances in heterostructure growth have enabled high-quality double quantum well systems~\cite{Tosato2022} and the first realizations of stacked quantum dots~\cite{Tidjani2023, Ivlev2024}, providing prospects for quantum dot arrays with connectivity beyond planar architectures. 

\begin{figure*}
    \centering
    \includegraphics[width=\linewidth]{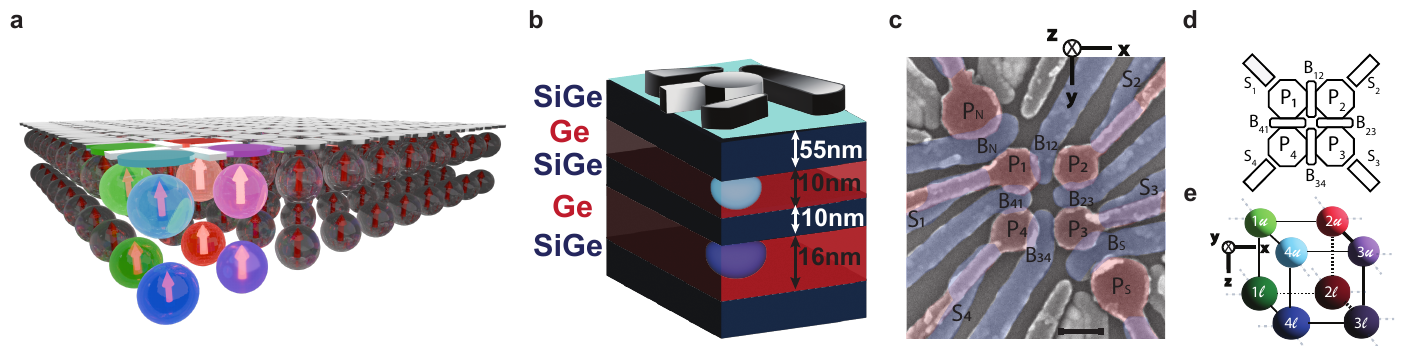}
    \caption{\textbf{Heterostucture and device layout for the three-dimensional quantum dot array a,} Schematic vision of how bilayer quantum heterostructures could be scaled to create bilayer qubit systems. The highlighted quantum dots forming a cube are the focus of the presented experimental work. \textbf{b,} Schematic of the Ge/SiGe double quantum well heterostructure, with 10\,nm of Si\textsubscript{0.2}Ge\textsubscript{0.8} separating two Ge quantum wells with a thickness of 10\,nm and 16\,nm.    
    \textbf{c,}  False-colored SEM image of a nominally identical device. A planar $2\times2$ array of plunger gates P\textsubscript{i} (red) is used to accumulate charges in the underlying germanium quantum wells. The barrier gates B\textsubscript{ij} (blue) and screening gates S\textsubscript{i} (blue) provide further control. 
    For charge readout, two RF-single hole transistors (SHT) are defined using the gates labeled P\textsubscript{N,S} and B\textsubscript{N,S}.  
    The scale bar denotes 100\,nm. 
    \textbf{d,}   Simplified gate layout showing the gates involved in triangulating the position of the quantum dots.  \textbf{e,} Schematic representation of the cuboid quantum dot configuration in this work, with $i$\textit{u}($\ell$) denoting a quantum dot in the upper (lower) quantum well, under plunger gate P\textsubscript{i}. }
  \label{fig:device_schematic}
    
\end{figure*}

Here, we explore the construction of quantum dots in three dimensions using multi-quantum well structures. In Fig.~\ref{fig:device_schematic}a, we conceptually illustrate how bilayer quantum wells could be used to realize intrinsically high connectivity quantum dot systems. Our device is based on a strained-germanium double quantum well embedded in a silicon-germanium heterostructure. We first tune a vertical double quantum dot, to then define quantum dot arrays on the facets of a cuboid. Understanding this system allows us to define eight quantum dots positioned at the corners of a cuboid. Finally, we demonstrate that the system supports quantum operation and enables the coherent displacement of single spins between quantum dots that reside in different quantum wells.

\section{The device}

Our experiments are performed on a silicon germanium (SiGe) heterostructure with two strained germanium quantum wells, depicted in Fig.~\ref{fig:device_schematic}b. The heterostructure is undoped and deposited on a Si$_{0.2}$Ge$_{0.8}$ strained relaxed buffer obtained by reverse grading on a Si(001) wafer \cite{Tosato2022, Sammak2019}. A SiGe spacer with a thickness of 55 nm separates the bilayer system from the gate stack. The top and bottom strained quantum wells have a thickness of 10\,nm and 16\,nm, and are separated by a 10\,nm Si\textsubscript{0.2}Ge\textsubscript{0.8} barrier. The quantum well separation is larger compared to earlier work that demonstrated vertical double quantum dots \cite{Tidjani2023, Ivlev2024}, aiming to reduce the tunnel coupling between the wells.

Control over the quantum dots is provided by a gate stack deposited on top of the heterostructure, as shown by a  false-colored SEM-image of a nominally identical device in Fig. \ref{fig:device_schematic}c. The device design is similar to a 2$\times$2 planar germanium qubit array including ohmics to define charge sensors at the corners \cite{Hendrickx2021}, but here the plunger gates are used to confine vertical double quantum dots, while barrier gates provide further control of the potential landscape. The device can therefore host up to eight quantum dots, with four quantum dots in each well, as schematically shown in Fig.~\ref{fig:device_schematic}e. Here, the spheres represent the upper (\textit{u}) or lower ($\ell$) quantum dots, under the plunger gates P\textsubscript{i} with $i=1-4$. Experimentally, we probe the position of the quantum dots in the array using triangulation, by sweeping the plunger gates against all surrounding gates and fitting the charge stability diagrams. In Fig.~\ref{fig:device_schematic}d, we show a simplified version of the gate layout only with the gates involved in the triangulation procedure. 

\section{Results}

\begin{figure*}
    \centering
    \includegraphics[width=\linewidth]{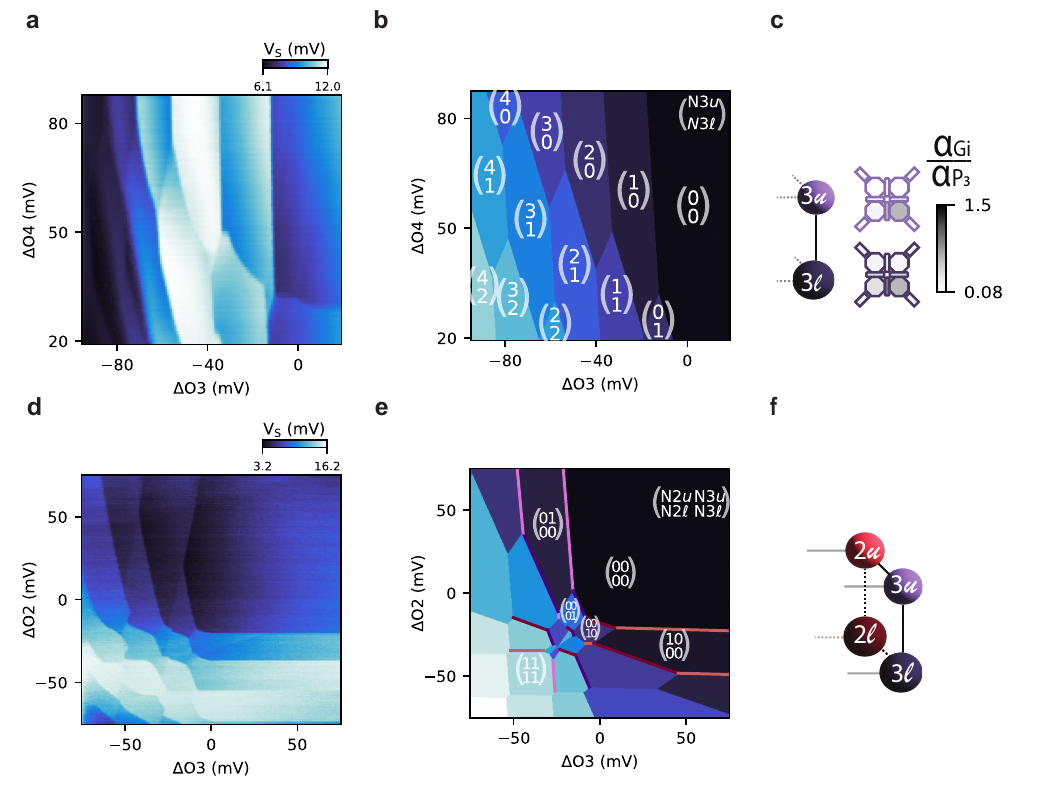}
    \caption{\textbf{Simulated and measured vertical double quantum dot charge stability diagrams in 10nm separated quantum wells.}
    \textbf{a,} Reflected signal of the rf-SHT underneath P\textsubscript{S} showing a charge-stability diagram that corresponds to a vertical double quantum dot located beneath P\textsubscript{3}, while P\textsubscript{4} is not yet accumulated.
    \textbf{b,} The simulated charge-stability diagram based on a constant-capacitance model~\cite{qarray} of a vertical double quantum dot localised under the plunger gate P\textsubscript{3}.
    \textbf{c,} Schematic of the respective vertical double dot and triangulation heatmap. The triangulation results stem form the fitted charge transition lines, providing the relative lever arm of each gate to the respective quantum dots. 
    \textbf{d,} Reflected signal of the rf-SHT underneath P\textsubscript{N} showing a charge-stability diagram that corresponds to a $2 \times 2$ array underneath P\textsubscript{2} and P\textsubscript{3} in the $yz$-plane. Two quantum dots are located underneath each plunger gate. The gates O2 and O3 represent the orthogonalized virtualized plunger gates, with respect to the quantum dots in the upper layer, QD2\textit{u} and QD3\textit{u}.
    \textbf{e,} Simulated charge stability diagram of a 2$\times$2 array aligned in the $xz$ plane. The capacitances are chosen to find qualitative agreement with the experimental results.
    \textbf{f,} Schematic denoting the respective facet of the quantum dot cuboid.}
    \label{fig2:simulate_exp_4dot}
\end{figure*}

Figure~\ref{fig2:simulate_exp_4dot} shows experimental and simulated charge stability diagrams of vertical double quantum dots. The simulations are performed using the Python package Qarray~\cite{qarray}, based on the constant-interaction model~\cite{VanDerWiel2003}. Vertical quantum dot arrangements give rise to distinct features as shown in Fig.~\ref{fig2:simulate_exp_4dot}a. In particular, we observe two reservoir transition lines strongly coupled to the same plunger gate and a long interdot transition line due to strong capacitative coupling between the vertically stacked quantum dots. 

To arrive in the regime of vertically stacked quantum dots in the few-hole regime, we tune the dc electrostatic potential of the dots independently. We do so by exploiting their difference in lever arms from the surrounding gates. See Appendix~\ref{supp_virtual_control} and Supp.~Videos for detailed experimental data.
In our simulations, we tune the capacitances to find reasonable agreement with the experimental data, see Figs.~\ref{fig2:simulate_exp_4dot}a and \ref{fig2:simulate_exp_4dot}b. The main difference is caused by latching which is present in the experiments and not considered in simulations. To provide further evidence of quantum dots residing in both wells, we perform triangulation based on the relative lever arms, see Fig.~\ref{fig2:simulate_exp_4dot}c and Appendix~\ref{supp_manualfitting}. For the pair P\textsubscript{i} - P\textsubscript{j}, we extract slopes by scanning P\textsubscript{i} and P\textsubscript{j} independently against all surrounding gates and extract the relative lever arm of the gates with respect to the plunger gate, at the same DC point. The charge stability diagrams are manually fitted to extract the slope of the charge transition lines in response to the surrounding gates, as shown in Appendix~\ref{supp_manualfitting}.

Identifying the charge states in stability diagrams with more than two quantum dots can be challenging, but we do find that bilayer systems can be of high-quality and stability. In Fig.~\ref{fig2:simulate_exp_4dot}e, the simulation shows two vertical double quantum dots. For low-hole fillings, we see a good correspondence between the experimental and simulated data of two vertical double quantum dots shown in Figs.~\ref{fig2:simulate_exp_4dot}d and \ref{fig2:simulate_exp_4dot}e.

\begin{figure*}[htbp]
    \centering
    \includegraphics[width=\textwidth]{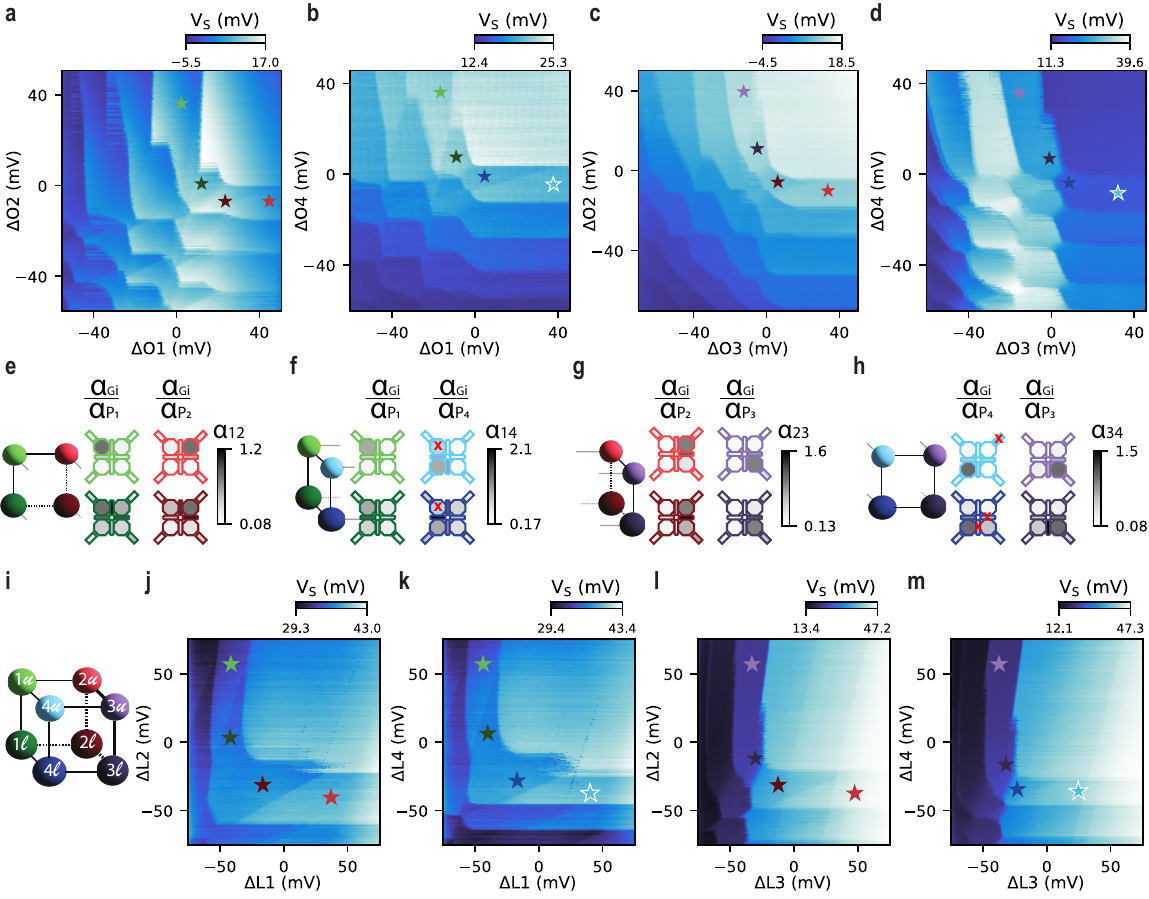}
    \caption{\textbf{Vertically aligned $\mathbf{2 \times 2}$ quantum dot arrays, and a $\mathbf{2 \times 2 \times 2}$ quantum dot array.}
    \textbf{a-d,}
    Reflectometry signal V\textsubscript{S} exhibiting stability diagrams of different facets of the cuboid. Here, the plunger gates that are not swept on the \textit{x} or \textit{y} axes are kept far from the accumulation voltage, so there are only four quantum dots in the system.
    \textbf{e-h,}
    Schematics and triangulation heatmaps of the quantum dots denoting the relative lever arms $\alpha_{\mathrm{G}_i}/\alpha_{\mathrm{P}_i}$ of the relevant gates.  e-h correspond to a-d in alphabetic order and the color code indicates the considered charging transition. See Fig.~\ref{fig:device_schematic}c for a labeled gate schematic. The relative lever arms are extracted from auxiliary gate-sweeps. 
\textbf{i}
Schematic of a cuboid quantum dot configuration.
\textbf{j-m,}
Stability diagrams corresponding to the eight quantum dot system. Here, the DC voltages (center of the 2D maps) lay $\sim$~10\,mV above single-hole accumulation. The color of the star in the single hole regime corresponds to the quantum dot in the schematic in \textbf{i}.
By sweeping pairs of virtualized plunger gate voltages $\Delta \mathrm{L}_i$, we identify eight distinct charging transitions that correspond to quantum dots on all corners of a cuboid underneath the plunger gates.
The virtual gate voltages $\Delta \mathrm{L}i$ correspond to orthogonal control of the lower quantum dots. The virtualization matrices are reported in the Appendix~\ref{supp:virtualisation}.}

    \label{fig:triangulation2x2sfigure}
\end{figure*}

\subsection{Four 2 x 2 Arrays}

Having established that we can tune between the planar and bilayer regime using the surrounding gates, we continue to form four $2\times 2$ arrays aligned in the \textit{xz} or \textit{yz} planes, across the different faces of the cuboid. In Fig.~\ref{fig:triangulation2x2sfigure}a-d we tune to have a $2\times 2$  array under the plunger gates P\textsubscript{1}-P\textsubscript{4}, P\textsubscript{1}-P\textsubscript{2}, P\textsubscript{3}-P\textsubscript{2}, and P\textsubscript{3}-P\textsubscript{4}, while keeping the other two plunger gates well above the accumulation voltage, in order to simplify quantum dot characterization. The $2\times 2$  array formed on the face of the device is indicated by the cartoon in Fig.\ref{fig:triangulation2x2sfigure}e-h, corresponding to the charge stability diagram above the cartoon.  Across all faces, the charge stability diagrams are rather similar, showing four distinct transition lines in each stability diagram, confirming the presence of four quantum dots. Importantly, we also observe four distinct transitions corresponding to the loading of a single hole, consistent with operating four quantum dots. 

To determine the location of the quantum dots, we repeat the triangulation procedure outlined previously. The colourmaps in Figs.~\ref{fig:triangulation2x2sfigure}e-h indicate the relative coupling of the gates to the respective plunger gates that are being swept. For the quantum dots in the top layer, the plunger gate always provides the strongest coupling to the quantum dot. The quantum dots associated with loading into the bottom layer exhibit a stronger overall coupling to the surrounding gates, consistent with being less electrostatically confined, in agreement with previous studies~\cite{Tidjani2023, Ivlev2024}. We note that we observe nonlinear capacitance, in particular for the barrier gates, such that the constant-interaction model can only be applied in limited regions. Additionally, we observe extended interdot lines~\cite{Yang2014}, in particular for the quantum dots under the gates P\textsubscript{2,4}, as they load holes via the quantum dots under P\textsubscript{1,3}.

Across all the triangulation diagrams, the relative lever arms indicate two quantum dots located predominantly under the same plunger gate. In some cases a barrier gate next to the relevant plunger gate has a lever arm larger than the plunger gate. We note that in our fabrication we pattern the barrier gates before the plunger gates, which increases the lever arms of the barrier gates.

\subsection{Quantum Dot Quboid}
We now extend our investigation to the simultaneous tuning of all 8 quantum dots, shown in Fig.~\ref{fig:triangulation2x2sfigure}i-m. We first tune a $2\times2$ array under the plunger gates P\textsubscript{1} and P\textsubscript{2}, after which we gradually accumulate under P\textsubscript{3} and P\textsubscript{4}. We find that the reduced confinement in this operational regime results in a larger overall tunnel coupling between the quantum dots. In particular, the stability diagrams indicate strong lateral coupling between the lower quantum dots. We see this most prominently in Fig.~\ref{fig:triangulation2x2sfigure}j,k. Nonetheless, we can still observe the latched loading lines of the lower quantum dots under P\textsubscript{2} and P\textsubscript{4}. Importantly, we find loading into four distinct charge states for the single-hole regime, indicating four quantum dots in each diagram. In these experiments, we keep the DC voltage point 15 mV above accumulation to avoid crossing loading lines from opposite quantum dots due to uncompensated crosstalk. The combination of these four charge stability diagrams together with the study on the individual facets of the cube show the loading of a single hole into eight quantum dots arranged in three dimensions.

\begin{figure}
    \centering
    \includegraphics[width=1.0\linewidth]{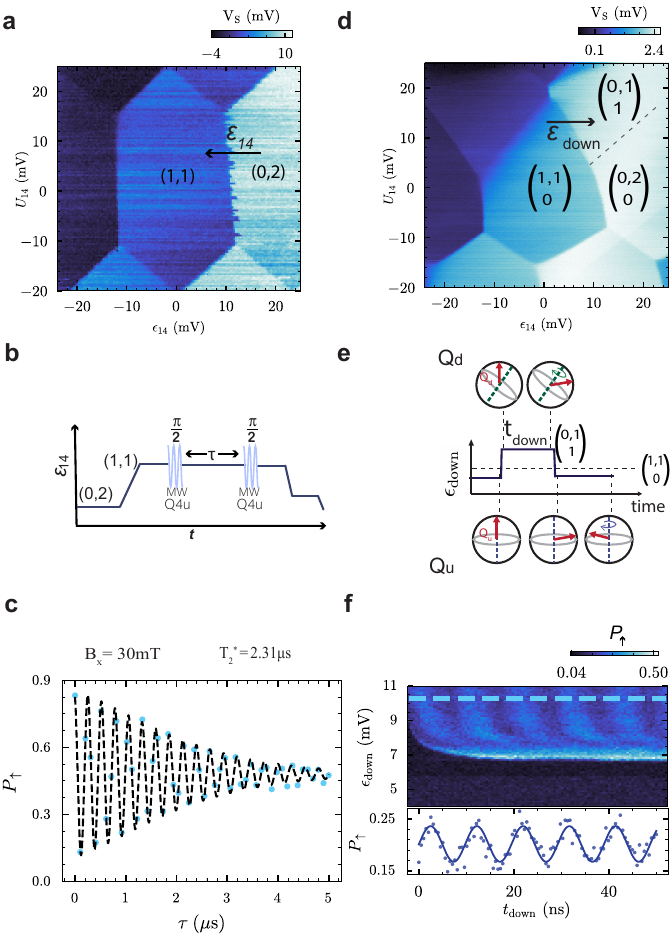}
    \caption{\textbf{Coherent operation in a bilayer heterostructure. a,} Isolated mode charge stability diagram of QD1\textit{u} and QD4\textit{u} when only the upper quantum well is accumulated with a total of 3 holes trapped in the system. 
    \textbf{b,} Ramsey sequence depiction on QD4\textit{u}, in the regime of \textbf{a}. 
    \textbf{c,} Experimental data of the ramsey sequence on QD4\textit{u}, from which we deduce a $T_2^* $ of 2.31$\mu$s at a magnetic field B\textsubscript{x}= 30\,mT. 
    \textbf{d,} Charge stability diagram with two upper and one lower quantum dot in isolated mode, with the charge state notation $\begin{psmallmatrix}
        QD1\textit{u} & QD4\textit{u}\\
        \hspace{17pt} QD\ell
    \end{psmallmatrix}$.
    \textbf{e} Depiction of the shuttling sequence between QD1\textit{u} and the lower quantum dot. A triplet is adiabatically initialised in QD4\textit{u} and QD1\textit{u}. The hole in QD1\textit{u} is then diabatically shuttled across $\epsilon_{\mathrm{down}}$ and allowed to precess for time $t_{\mathrm{down}}$, before shuttling back to the original quantum dot and read out by Pauli spin blockade between two quantum dots in the upper layer.. 
    \textbf{f,} Shuttling induced oscillations by diabatically pulsing across $\epsilon_{down}$ into the lower quantum dot and back to the original quantum dot, as detailed in \textbf{e}. The lower panel shows a linecut of the oscillations on the dotted line in the upper panel at $\epsilon_{\mathrm{down}}= 10.27$\,mV.}
    \label{fig:coherentop}

\end{figure}

\subsection{Coherent spin operation in a bilayer system}

A key question for the prospects of multilayer quantum wells for quantum information is whether spin coherence is compromised in multi-well systems. As a first step to address this question, we use the pair QD1\textit{u} and QD4\textit{u} and tune to the few-hole regime, see Fig. \ref{fig:coherentop}a. We initialise a triplet $\ket{\downarrow, \downarrow}$ state by adiabatically ramping from the (2,0) to the (1,1) charge state. We then perform a spin rotation through resonant microwave control and readout the resulting spin states using Pauli spin blockade. Fig.~\ref{fig:coherentop}b shows Ramsey experiments using this method and charge readout using the sensor dot. We measure $T\textsubscript{2}^*= 2.31$\,\textmu s at $B_x = 30$ mT. This value is appreciable given the typical timescales required to execute quantum gates with germanium qubits and falls within the range of measured coherence times for hole spins in single strained germanium quantum wells \cite{John2024}, suggesting that the presence of multiple quantum wells does not significantly affect the quantum coherence and may allow for high-fidelity operation.

Another intriguing question is whether hole spins can be shuttled between the quantum wells through the silicon-germanium barrier. To study this, we choose to operate in an isolated mode (see Appendix~\ref{app_isolated} for details and stability diagrams in isolated mode) \cite{weehan2x2}. This is achieved by reducing the tunnel coupling of the quantum dots to the reservoir during operation, see Fig. \ref{fig:coherentop}d. To test whether we can coherently shuttle between the two quantum wells, we utilize the large sensitivity of the spin quantization axis of hole spins in germanium to the electric and magnetic fields and implement a shuttling protocol \cite{vanriggelendoelman2023coherent, Wang2024}. Figs.~\ref{fig:coherentop}e and \ref{fig:coherentop}f show the protocol and experimental results. In particular we find that once we shuttle from the top to the bottom quantum well, the larmor precession is evolving around a tilted axis, resulting in qubit rotations. We note that the finite visibility observed is expected to be limited by the difference between quantization axes. Further research is needed to investigate the quality of vertical hopping and to which extent the difference in the g-tensor can be engineered, for example by differences in the strain between the quantum wells.




\section{Discussion and outlook}

Silicon-germanium heterostructures define a versatile and high-quality platform. We have investigated quantum dot operation in multilayer quantum wells and demonstrated that quantum dots can be interconnected in three spatial dimensions.
We find that three-dimensional quantum dot structures yield stable and clean charge-stability diagrams and allow for coherent spin control.

Operating three-dimensional quantum dot arrays may offer new opportunities for quantum simulation and computation~\cite{farina2025site}. In our work, we have also identified several challenges in building such systems. Compared to planar systems, multilayer systems have smaller operational windows due to a low number of control parameters compared to the system size. Future designs may address this by the inclusion of additional gates, where the small effective mass of germanium can be beneficial~\cite{Lodari2019, Scappucci2020TheRoute}, as it allows increasing the quantum dot size. Furthermore, the simultaneous confinement of holes in the two quantum wells and operation in the few-hole regime depend strongly on heterostructure design choices, as well as on aspects such as quantum dot size and applied gate voltage. This highlights the need for a co-design approach involving material growth, device fabrication, and system operation. Importantly, the observed quality of the double quantum well system offers promising prospects for model-based design and further advancement.

In our experiments, we qualitatively find that quantum dots in multilayer systems can be strongly coupled. This would be beneficial for quantum simulations where large couplings are desired. Future experiments may also investigate strained quantum wells with larger separation to reduce the coupling between the layers to optimize high-connectivity quantum dot arrays for quantum computing.

\newpage
\section{Acknowledgements}
We would like to acknowledge useful discussions with C. D\'eprez, A. Ivlev, F. Borsoi, D. Jirovec and I. Fern\'andez de Fuentes.

\section{Data Availability}

The raw data and analysis supporting the findings of this study are openly available in a Zenodo repository:
https://zenodo.org/records/16305716.

\section{Funding}
We acknowledge support through an NWO ENW grant and an ERC Starting Grant QUIST (850641). 
This research was supported by the European Union through the Horizon 2020 research and innovation program under grant agreement No. 101069515 (IGNITE). This research was sponsored in part by the Army Research Office (ARO) under Award No. W911NF-23-1-0110 and by The Netherlands Ministry of Defence under Awards No.QuBits R23/009. The views, conclusions, and recommendations contained in this document are those of the authors and are not necessarily endorsed nor should they be interpreted as representing the official policies, either expressed or implied, of the Army Research Office (ARO) or the U.S. Government, or The Netherlands Ministry of Defence. The U.S. Government and The Netherlands Ministry of Defence are authorized to reproduce and distribute reprints for Government purposes notwithstanding any copyright notation herein.

\section{Competing Interests}
M.V. and G.S. are founding advisors of Groove Quantum BV and declare equity interests. The remaining authors declare that they have no competing interests.

\section{Author contributions}

H.T, D.D. M.C. and J.H.U conducted the experiments. H.T, and D.D. performed the analysis. M.C. and B.v.S performed the simulations. H.T. designed the device, S.O. fabricated the device, L.E.A.S. and G.S. supplied the heterostructures. H.T, D.D., J.H.U, M.C. and M.V. wrote the manuscript with input from all authors. M.V. supervised the project.


\section{Methods}
\subsection{Device fabrication}
The device is fabricated on a Si\textsubscript{$x$}Ge\textsubscript{$1-x$}/Ge/Si\textsubscript{$x$}Ge\textsubscript{$1-x$}/Ge/Si\textsubscript{$x$}Ge\textsubscript{$1-x$} heterostructure, where $x = 0.2$, grown by reduced pressure chemical vapour deposition. The virtual substrate upon which the heterostructure is grown consists of a silicon substrate, upon which there is a \SI{1.6}{\micro m} relaxed Ge layer; a \SI{1}{\micro m} graded Si\textsubscript{$x$}Ge\textsubscript{$1-x$} layer, with a final Ge composition of x = 0.2.  On top of the SiGe virtual substrate, the bilayer system comprises in ascending order a \SI{16}{nm} thick bottom Ge quantum well, a \SI{10}{nm} thick SiGe barrier, a \SI{10}{nm} thick top Ge quantum well, and a \SI{55}{nm} thick SiGe spacer. At the top of the stack, a sacrificial Si cap is grown to provide a native SiOx layer. We define ohmic contacts using electron beam lithography and remove the Si cap in the exposed area using a buffered oxide etch. We then evaporate a \SI{30}{nm} platinum layer and contact the quantum wells using a 10-minute rapid thermal anneal at $400^\circ$C, forming platinum germanosilicides. The ohmic layer is isolated using a \SI{7}{nm} layer of Al\textsubscript{2}O\textsubscript{3} grown by atomic layer deposition. Electrostatic gates used to define the quantum dots are patterned in two layers (3/\SI{17}{nm} and 3/\SI{37}{nm} of Ti/Pd.) and are separated by a \SI{5}{nm} layer of Al\textsubscript{2}O\textsubscript{3}.

Devices are screened at \SI{4}{K} in liquid helium. Experiments reported in this paper are carried out in a Bluefors LD400 dilution refrigerator with a base temperature of \SI{10}{mK}. An in-house built battery-powered SPI rack (see \url{https://qtwork.tudelft.nl/~mtiggelman/spi-rack/chassis.html}) is used to set direct-current (DC) voltages, and a Qblox Cluster arbitrary waveform generator (AWG) supplies alternating-current (AC) pulses via coaxial lines. The DC and AC voltages are combined on the printed circuit board (PCB) with bias-tees before routing to the gates. For reflectometry measurements the DC bias across the sensor dots is set to 0\,V. Further details on the measurement procedure can be found in \cite{Wang2024}. Magnetic fields are generated by an American Magnetics vector magnet, with maximum field strength of 1\,T in arbitrary directions.


%

\onecolumngrid

\begin{appendices}
\onecolumngrid

\setcounter{table}{0}
\setcounter{figure}{0}

\setcounter{section}{0}
\section{Sensor Tuning}

\begin{figure}[H]
    \centering
    \includegraphics[]{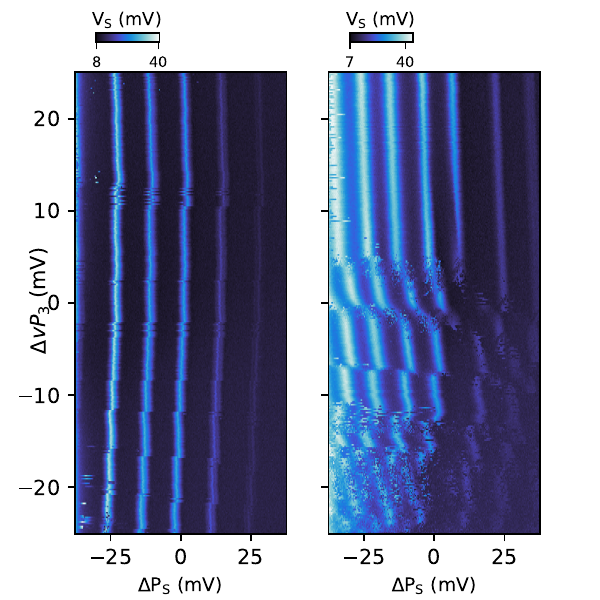}
    \caption{\textbf{Sensor operation.} In the left panel, the sensor is a single quantum dot. The experiments show measurement of a quantum dot under P\textsubscript{3}. In the right panel, the sensor contains a vertical double quantum dot. } 
    \label{Suppfig:sensingdot}
\end{figure}
As the sensing dots can also be composed of a vertical double quantum dot, this can interfere with the charge signal when detecting charge transitions of the quantum dots in the plunger array. To avoid this, we strongly confine the sensing dot in order to have a single quantum dot. This can come at the expense of reduced contrast in the charge transitions of the quantum dots in the array. A vertical double quantum dot in the sensing dot may compromise the quality of the rf-signal, due to the crosstalk within the sensing dot, interfering with the charge signal from the device. 

\section{Virtual control: Tuning between the single layer and the bilayer regime}
\label{supp_virtual_control}
\begin{figure}[H]
    \centering
    \includegraphics[width=\linewidth]{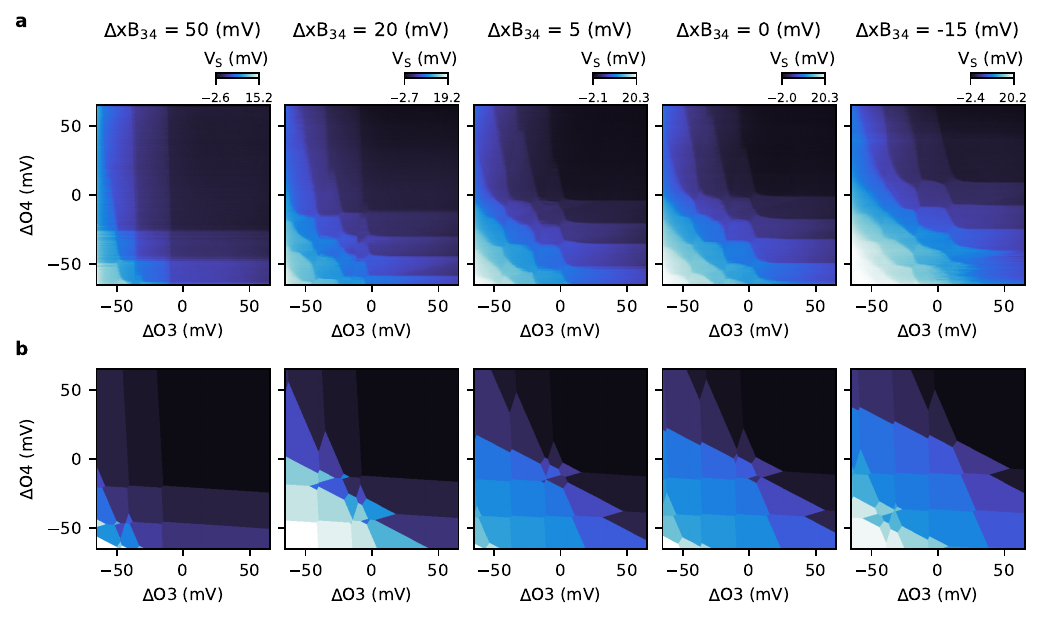}
    \caption{\textbf{Tuning between the planar regime and the bilayer regime.} \textbf{a} We can actively tune between the single layer regime, seen in the left most panel, to the bilayer regime (right most panel). We observe four charge transition lines, with two dots strongly coupled to O3, and two dots strongly coupled to O4, indicating the presence of two vertical double dots in the system. \textbf{b} We qualitatively replicate this effect using quantum capacitance simulator Qarray\cite{qarray}. Deviations between experiment in \textbf{a} and simulation \textbf{b} is attributed to change in the quantum dot shape, and thereby leverarms, as function of the barrier gate which is not captured in the constant capacitance model.}
    \label{Suppfig:singletobilayerregime_B34}                       
\end{figure}

\begin{figure}[H]
    \centering
    \includegraphics[width=\linewidth]{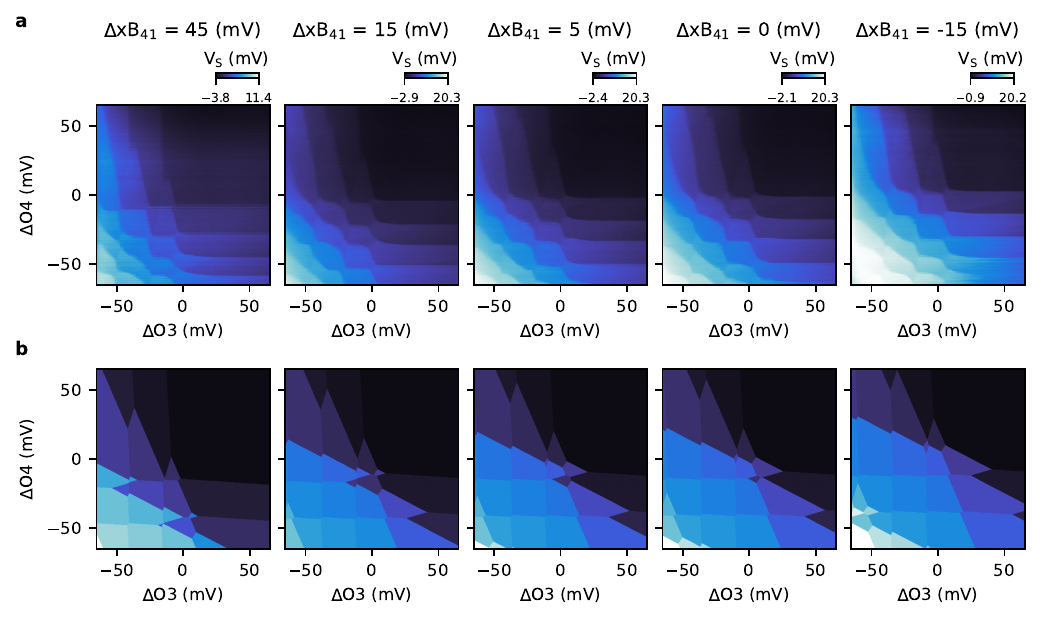}
    \caption{\textbf{Tuning between single and bilayer regime for a single vertical double quantum dot.}
    \textbf{a} By pulsing on xB\textsubscript{41}, we are able to tune from a vertical double quantum dot underneath P\textsubscript{4} (left most plot) to just a single quantum dot in the upper layer (right most plot). The vertical double quantum dot underneath P\textsubscript{3} is only slightly affected. \textbf{b} Qarray simulations using the same capacitance parameters as in Fig. \ref{Suppfig:singletobilayerregime_B41}b. Qualitative features and behavior is replicated in the simulation.}
    \label{Suppfig:singletobilayerregime_B41} 
\end{figure}

The lower layer quantum dots can be made energetically favorable by modifying the potential landscape using the surrounding barrier, plunger (if unaccumulated) and screening gates.
Examples with experimental data and simulations are shown in Figure \ref{Suppfig:singletobilayerregime_B34} and \ref{Suppfig:singletobilayerregime_B41} for the system in the configuration of figure \ref{fig:triangulation2x2sfigure}h. 

In Fig.\ref{Suppfig:singletobilayerregime_B34} we tune the device completely from the bilayer regime, containing 4 quantum dots in total; 2 dots in the upper and lower layers beneath gate P\textsubscript{3} and P\textsubscript{4}, to the single layer regime, containing just only the two quantum dots in the upper layer.
This is achieved by a positive voltage pulse on xB\textsubscript{34}, the barrier between the two relevant plungers, to increase the confinement of the quantum dots.
This tunes the chemical potential of the lower quantum dots away with respect to the upper quantum dots. 
In the leftmost panels of Fig. \ref{Suppfig:singletobilayerregime_B34}a,b we have reached the planar regime for the first hole filling, as is evident from the two charge loading lines at the first filling. 
We do note that charge loading lines of the lower quantum dots appear at higher occupation at more negative O3 and O4 gate values.
Conversely, when pulsing more negatively on xB\textsubscript{34} (going left-right), we tune the lower quantum dots to be more energetically favorable to occupy. 
This is seen as moving the loading lines of the lower quantum dots towards more positive  O3 and O4 gate voltages, towards the first loading. 
After this point, pulsing more negatively on the barrier increases the charge regions containing any holes in the lower layer. 
This is observed for example in the right most plot, where in the first hole filling the charge region in which the hole is in either 
QD3$\ell$ or QD4$\ell$ has enlarged.
The experimental data are qualitatively replicated using Qarray, as depicted in Fig. \ref{Suppfig:singletobilayerregime_B34}b. 
The key observations from the experiment are seen in the simulation as well. 

Alternatively, we can also pulse on xB\textsubscript{41} as shown in Fig.\ref{Suppfig:singletobilayerregime_B41}.
Barrier gate B\textsubscript{41} is located closer to plunger P\textsubscript{4} and further away from P\textsubscript{3}.
As such, we expect a larger leverarm of this barrier gate to QD4$\ell$ compared to QD3$\ell$. 
This is indeed seen in Fig.\ref{Suppfig:singletobilayerregime_B41}, where we actively tune from a vertical double quantum dot (right most panel) to just a single quantum dot (left most panel) in the upper layer, underneath P\textsubscript{4}. 
Again, we note that the first loading of QD4$\ell$ in the left most panel is still observed, albeit at more negative O4 gate value. 
In contrast, the vertical double quantum dot underneath P\textsubscript{3} is barely affected by the pulsing on xB\textsubscript{41}. 
It indicates that we are able to detune the chemical potential of QD4$\ell$ more effective than QD3$\ell$ using B\textsubscript{41}. 
This demonstrate the ability to control the occupation of quantum dots in the upper and lower layer independently. 
Using the same capacitance input for the Qarray simulation in Fig.\ref{Suppfig:singletobilayerregime_B34}, we are able to reproduce this behaviour in the simulations shown in Fig.\ref{Suppfig:singletobilayerregime_B41}b.

Additional examples where we tune the device between the single and bilayer regime using other surrounding gates, such as screening and unaccumulated plunger gates, are found in supp videos.

\section{Multi-Dot isolated mode}
\label{app_isolated}
\begin{figure}[H]
    \centering
    \includegraphics[width=\linewidth]{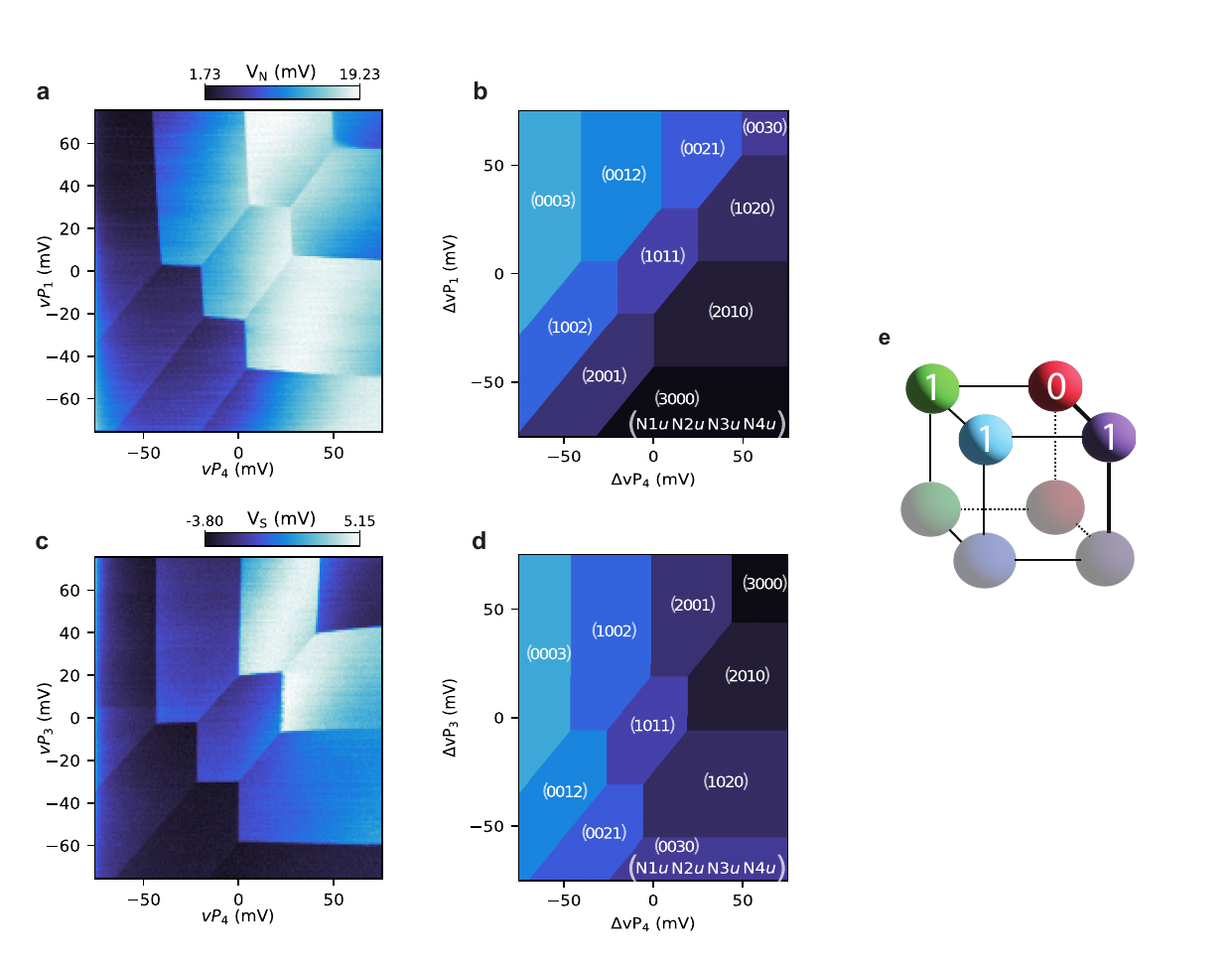}
    \caption{
    \textbf{
    Triple dot charge stability diagrams in isolated mode while only the upper layer is accessible}. 
    Charge stability diagrams for the QD1\textit{u}-QD4\textit{u} pair \textbf{(a)} and QD3\textit{u}-QD4\textit{u} \textbf{(c)} pair considering a total of 3 holes and 4 QDs. \textbf{b,d} Simulated charge stability diagrams for a triple dot system hosting three holes, corresponding to \textbf{a,c} respectively. \textbf{a,c} are acquired at the same DC point and pulsing $\pm$100 mV on the virtualized plungers. \textbf{e} Schematic representation of cuboid quantum dot configuration. The number in each sphere represent the charge occupation in the corresponding dot in isolated mode at the DC point at the center of the charge stability diagram.
    }
    \label{fig:isolated_planar}
\end{figure}

Demonstrating the capability to control the charge degree of freedom of quantum dots formed in a bilayer Ge/SiGe heterostructure is a key requirement for spin operation. Generally, precise control over tunnel coupling between quantum dots is required to achieve high-quality initialization, operation, and readout for both ST and LD qubits.
The barrier between the sensor and the first dot (B\textsubscript{S} and B\textsubscript{N}) in the array affects the charge loading of the upper or lower quantum dots from the sensor reservoir differently, leading to asymmetric charge decay rates for the upper- or lower-layer quantum dots. Moreover, when the lower dot is accumulated, the (1,1) and (2,0)/(0,2) charge states are no longer connected. This prevents the initialization of spins in the upper layer and the subsequent shuttling of single spins down without implementing pulsing schemes for initialization and readout.
For this reason, the system is operated in the closed regime, also known as the isolated regime, where a fixed number of holes is trapped\cite{weehan2x2}.  
This is achieved by pinching off the barriers that separate the sensors from the central dot array, namely B\textsubscript{S} and B\textsubscript{N}. 

An example of a triple dot charge stability diagram for the upper layer in isolated mode is shown in Fig. \ref{fig:isolated_planar}. The holes are first loaded via P\textsubscript{S}, then by increasing the barrier voltage (B\textsubscript{S}) below the accumulation threshold tunneling to the ohmic reservoir is suppressed. All the lines in the charge stability diagram represent interdot transitions which preserve the total number of charges.
Three holes are trapped in the system between QD1\textit{u}, QD3\textit{u} and QD4\textit{u}. Each QD hosts 1 hole at the DC voltage point. Fig. \ref{fig:isolated_planar}a shows the P\textsubscript{1}-P\textsubscript{4} gate pair and \ref{fig:isolated_planar}c shows P\textsubscript{3}-P\textsubscript{4}. The resulting charge stability maps are consistent with the simulation in isolated mode performed using Qarray as shown in Fig.\ref{fig:isolated_planar}b, \ref{fig:isolated_planar}d. The labeling of the charge state is independently assigned using the simulation.

\begin{figure}[H]
    \centering
    \includegraphics[width=\linewidth]{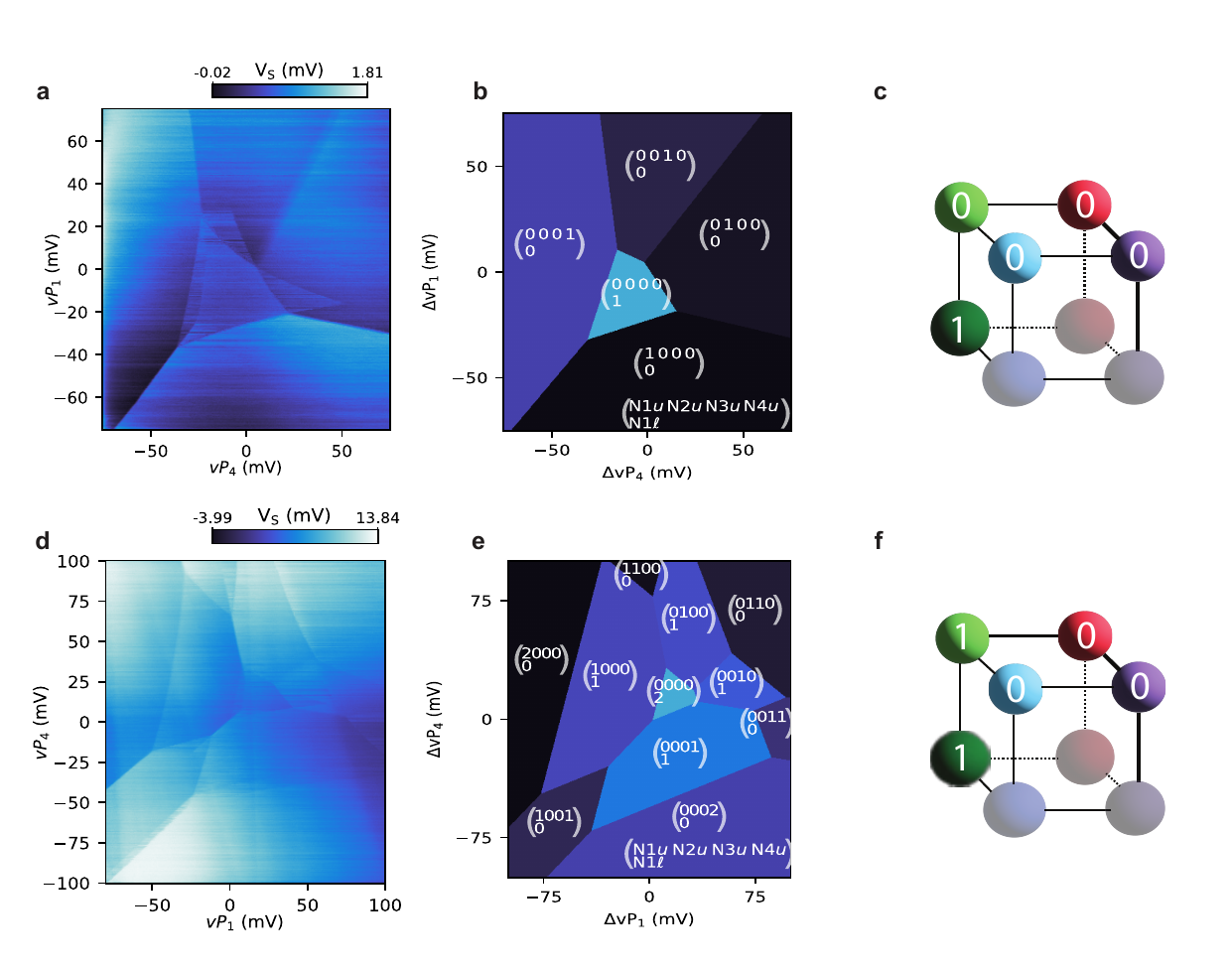}
    \caption{\textbf{Multi-dot charge stability diagrams isolated mode with the lower layer occupied}. \textbf{a, c} Charge stability diagram, scanning P\textsubscript{1} and P\textsubscript{4}, of a 5 quantum dot array in isolated mode hosting 1 and 2 holes respectively. The isolated hole(s) has access to 1 lower layer quantum dot (QD1$\ell$) and all the four upper layers dots.\textbf{b,d} Simulated charge stability diagrams corresponding respectively to \textbf{a, c}. Both simulations are made with the same capacitance matrices, chosen to qualitatively fit the experiment best. \textbf{c,f} Schematic representation of cuboid quantum dot configuration. The number in each sphere represents the charge occupation of the corresponding dot in isolated mode at the DC point at the center of the charge stability diagram.}
    \label{fig:isolated_vertical_app}
\end{figure}

By modeling the system in isolated mode, we can also study how the presence of the lower quantum dots will affect the shape of the isolated charge stability diagram. 
We note in Fig. \ref{fig:isolated_vertical_app} that the presence of an extra dot in the lower layer appears as an additional set of transition lines orthogonal to the interdot line between the pair of upper quantum dots from which the plungers are swept.

\begin{figure}
    \centering
    \includegraphics[width=\linewidth]{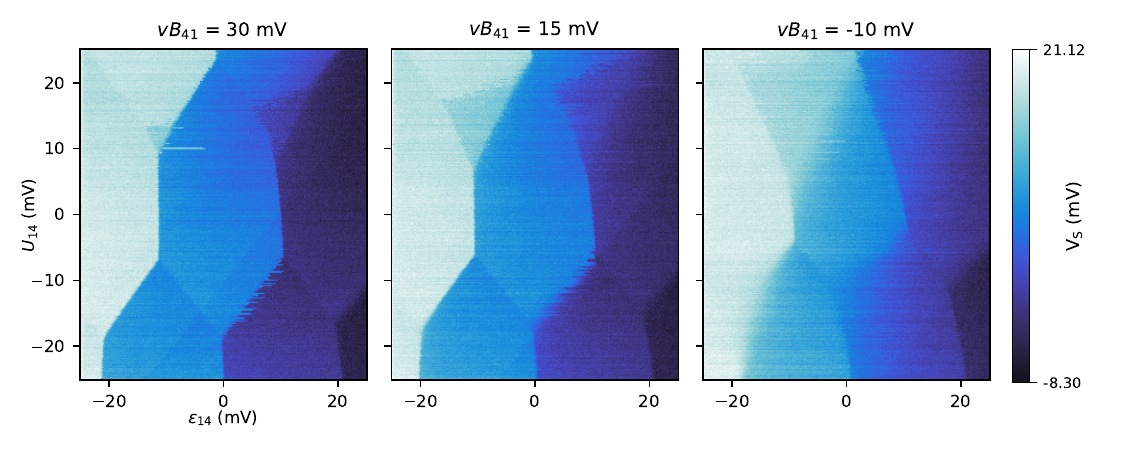}
    \caption{\textbf{Series of CSD in isolated mode which shows the lower layer tunability using the barrier vB\textsubscript{41}.}}
    \label{fig:isolated_tunable}
\end{figure}

Indeed, by decreasing the barriers and screening of the 2x2 quantum dot array it is possible to bring one the lower dots close to accumulation. An example of this is shown in Fig.\ref{fig:isolated_tunable}, where the barrier between the P\textsubscript{1} and P\textsubscript{4} dots (vB\textsubscript{41}) is pulsed in steps of 20 mV. An additional charge state appears above the (2,0,1)-(1,1,1) interdot transition as well as (0,2,1)-(1,1,1) transition.
A comparison between simulation and experiment for a total charge occupation of 1 and 2 holes with an additional lower layer quantum dot in isolated mode is shown in Fig.\ref{fig:isolated_vertical_app}.
The tilt of the up-to-low interdot line is related to how strong the plunger-to-lower layer lever arm is.
Interesting to note in Fig.\ref{fig:isolated_tunable} is that also the latching of the up-to-down transition depends on the barrier value which points towards a tunable up-to-down interdot coupling. The ability to operate both layers without any changes in hole occupation throughout initialization, control and readout is a main prerequisite for scaling it up to large three-dimensional arrays.

\section{Virtualisation}
\label{supp:virtualisation}
We use an embedded form of virtual gate matrices. This gives us the flexibility easily to define the same gate in multiple different ways. We begin by virtualising the gates to the sensing dots, then gate-gate. Here we show the inverted virtual gate matrix  $M^{-1}$, which is defined as $M^{-1} * \text{Virtual Gates}= \text{Real Gates}$.
The virtualisation may change between different tuning regimes, and the exact forms can be found in the metadata.

\begin{figure}[H]
    \centering\includegraphics[width =\linewidth]{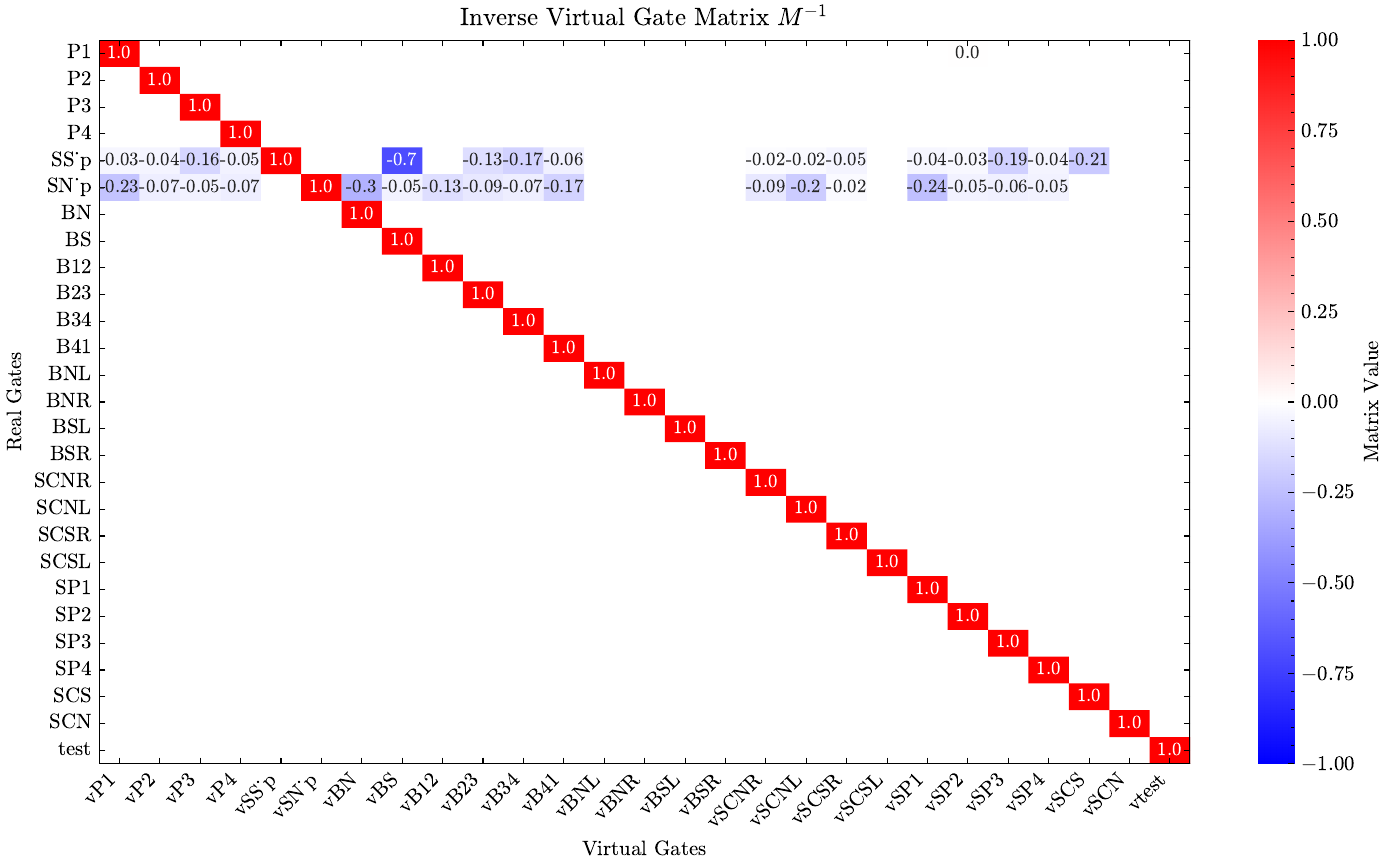}
    \caption{Virtual gate matrix for compensating the sensor signal.}
    \label{}
\end{figure}
\begin{figure}[H]
    \centering
    \includegraphics[width =0.9\linewidth]{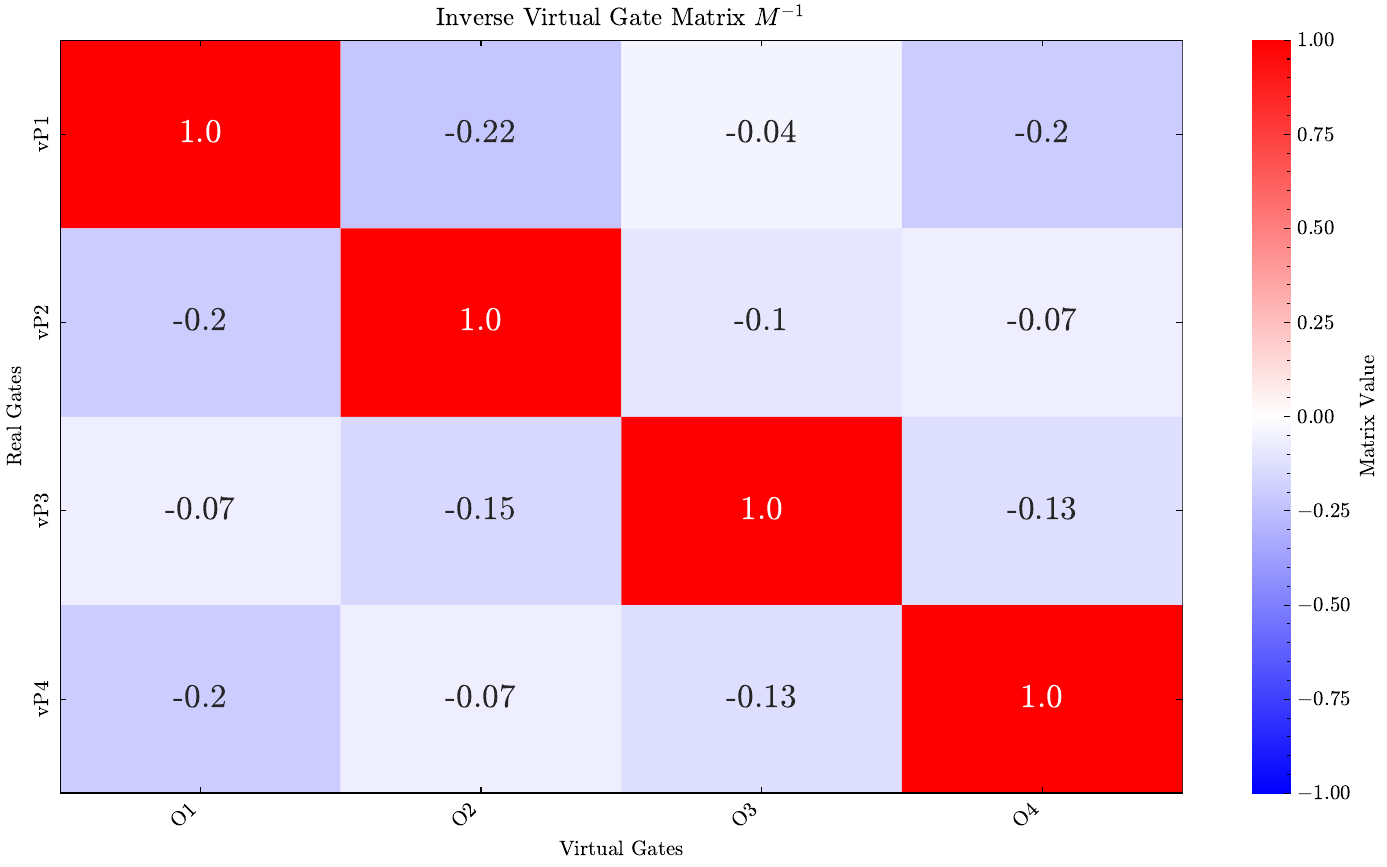}
    \caption{Plunger to Plunger gate virtualisation for the upper quantum dots.}
    \label{}
\end{figure}
\begin{figure}[H]
    \centering
    \includegraphics[width =0.9\linewidth]{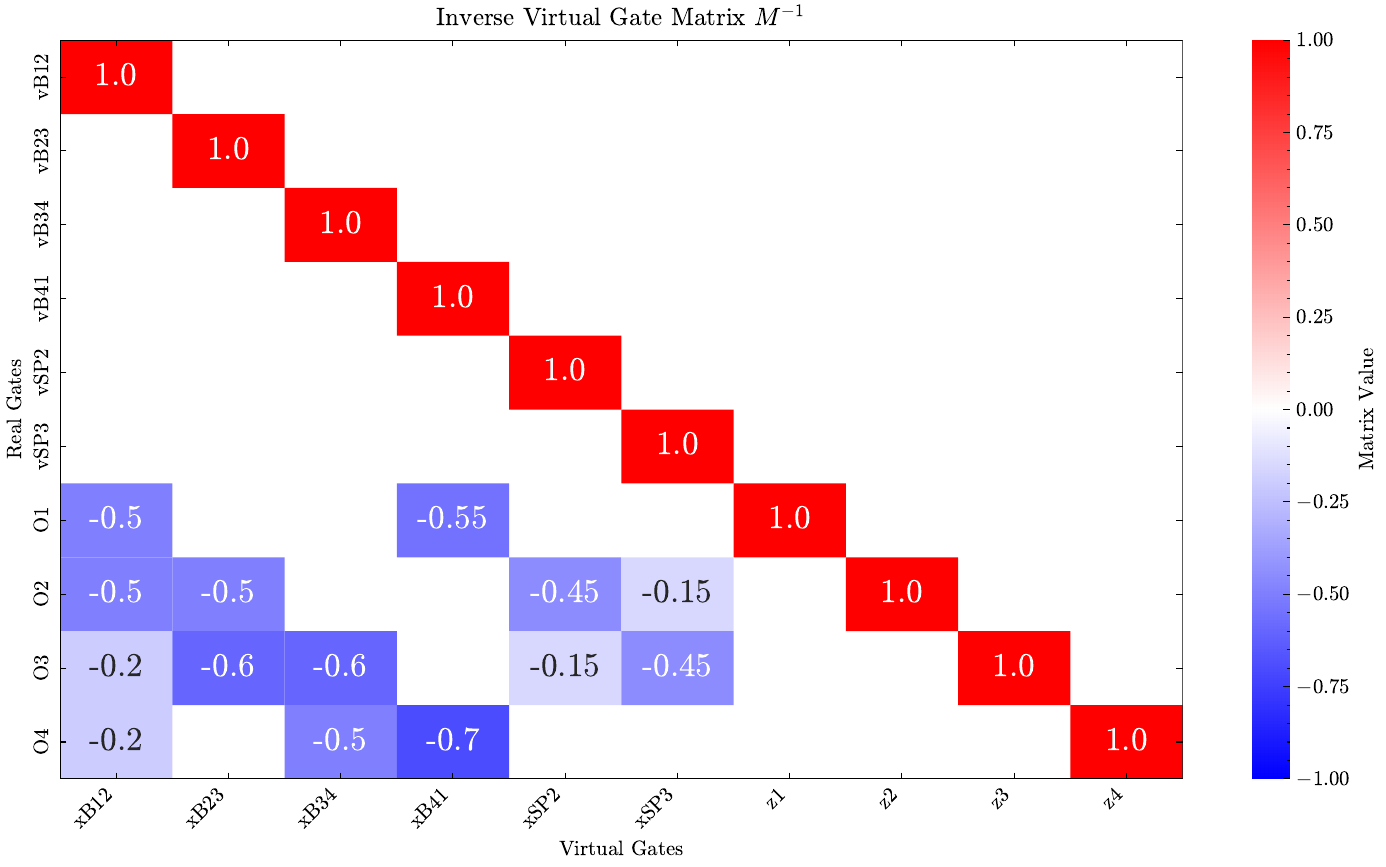}
    \caption{Barrier to Plunger gate virtualisation.}
    \label{}
\end{figure}

\begin{figure}[H]
    \centering
    \includegraphics[width =0.9\linewidth]{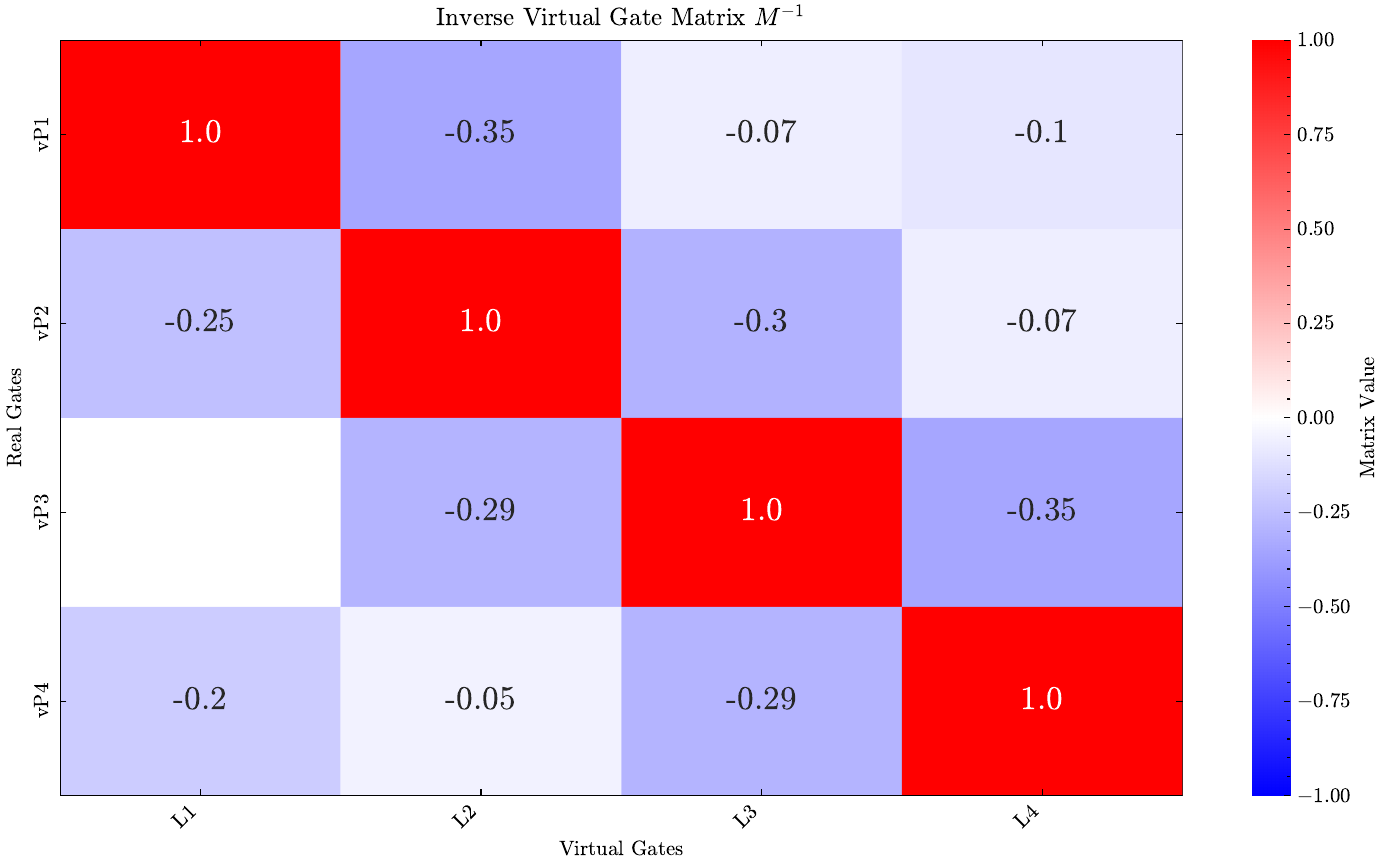}
    \caption{Plunger gate to Lower quantum dot virtualisation matrix. }
    \label{}
\end{figure}


\begin{figure}[H]
    \centering
    \includegraphics[width =0.9\linewidth]{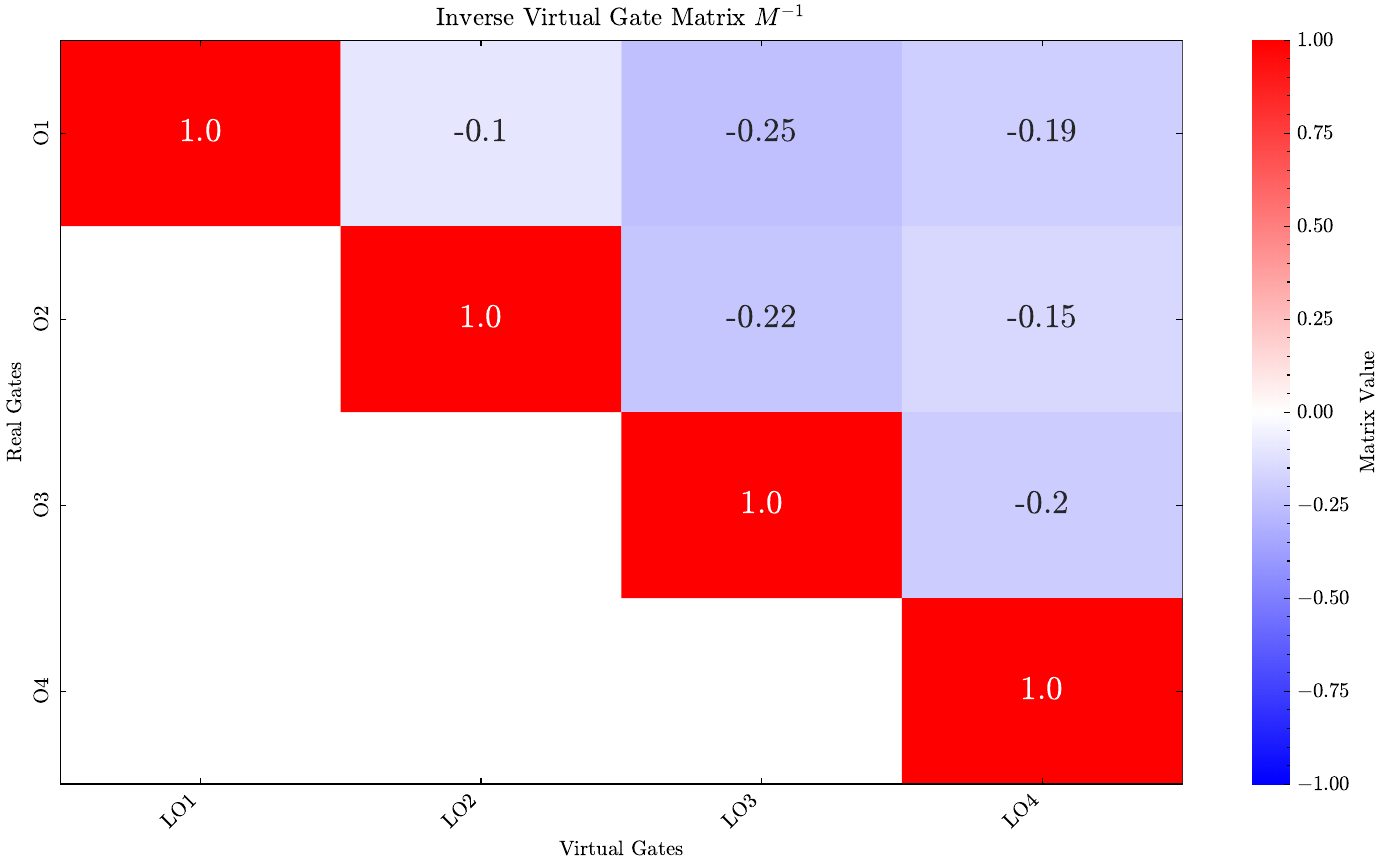}
    \caption{Orthogonalised plunger gate virtualisation to lower orthogonal gates.}
    \label{}
\end{figure}

\begin{figure}
     \centering
     \includegraphics[width =0.8\linewidth]{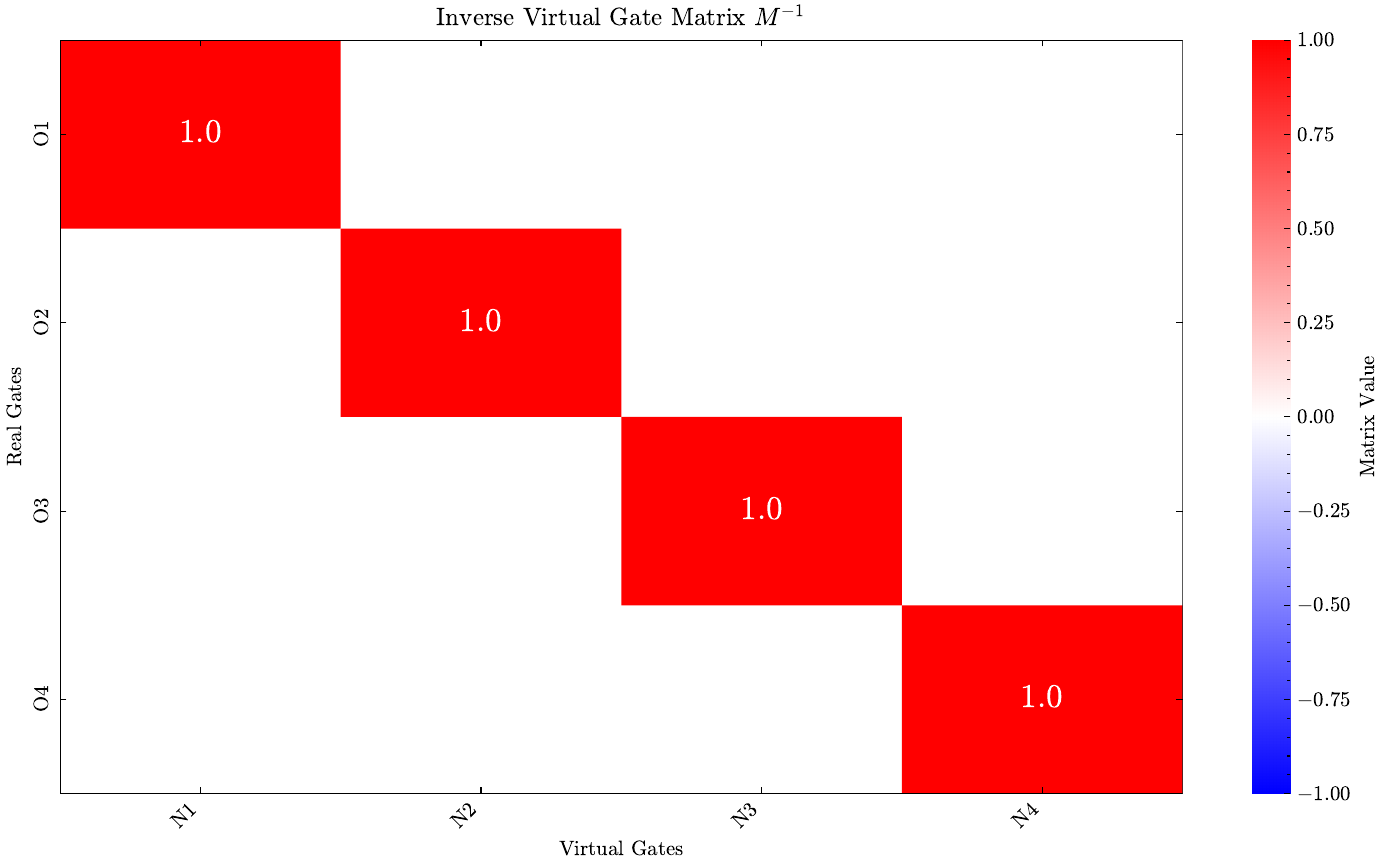}
     \caption{One to one orthogonal gates to normalised gates mapping.}
     \label{fig:enter-label}
\end{figure}

\begin{figure}[H]
    \centering
    \includegraphics[width =0.8\linewidth]{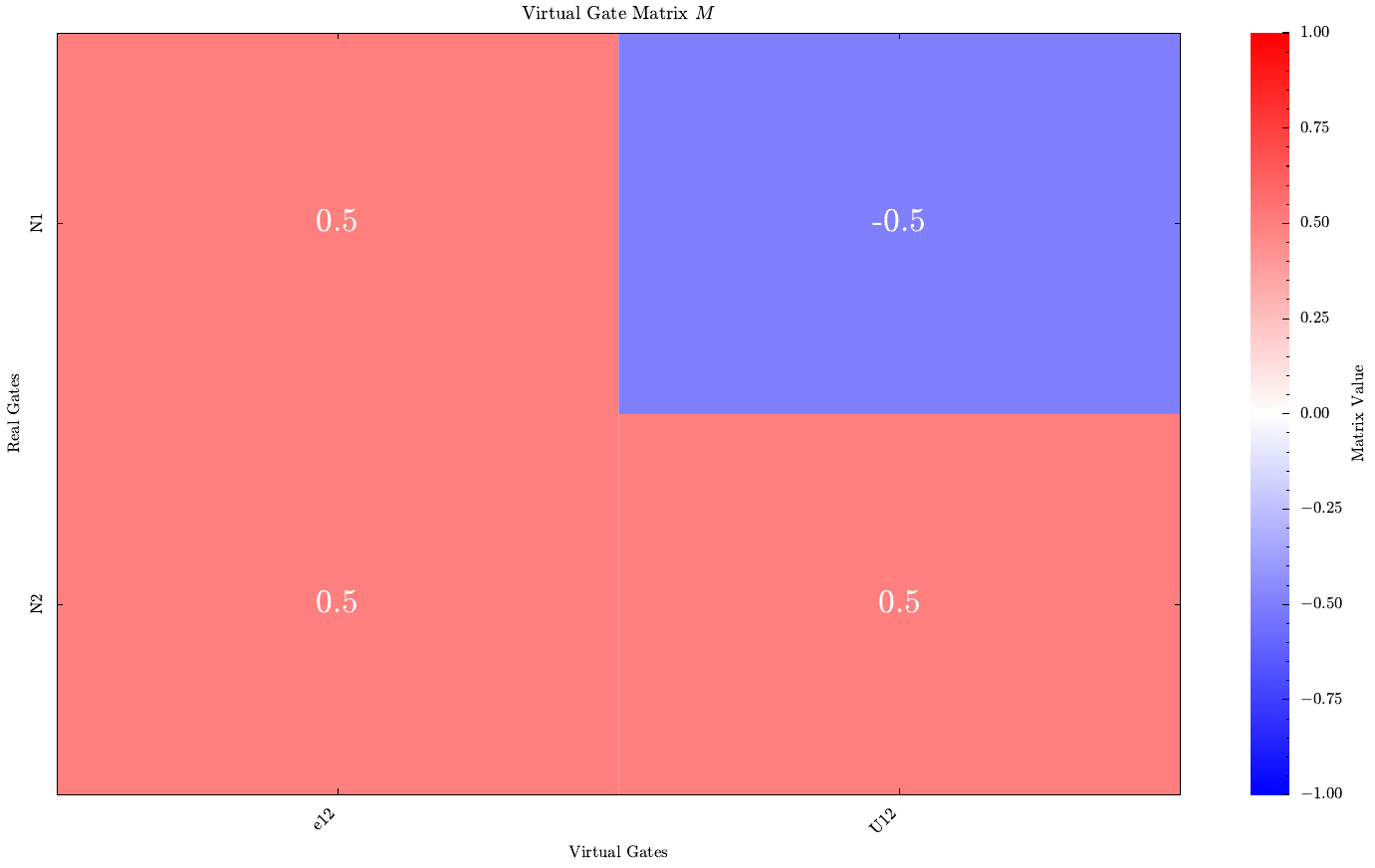}
    \caption{Detuning $\epsilon_{12}$ and Potential energy $U_{12}$ virtual gate matrix.}
    \label{}
\end{figure}

\begin{figure}[H]
    \centering
    \includegraphics[width =0.8\linewidth]{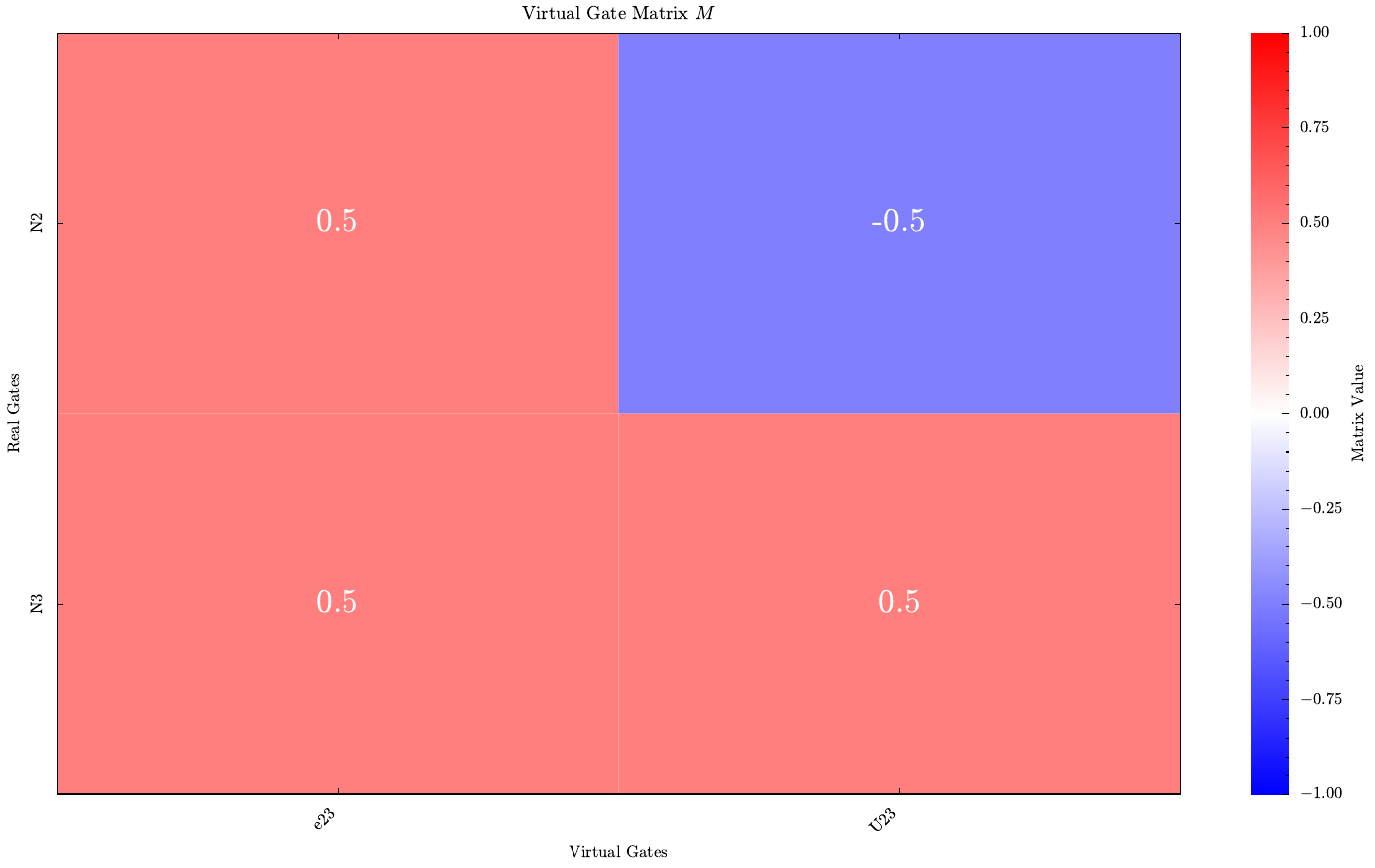}
    \caption{Detuning $\epsilon_{23}$ and Potential energy $U_{23}$ virtual gate matrix.}
    \label{}
\end{figure}

\begin{figure}[H]
    \centering
    \includegraphics[width =0.8\linewidth]{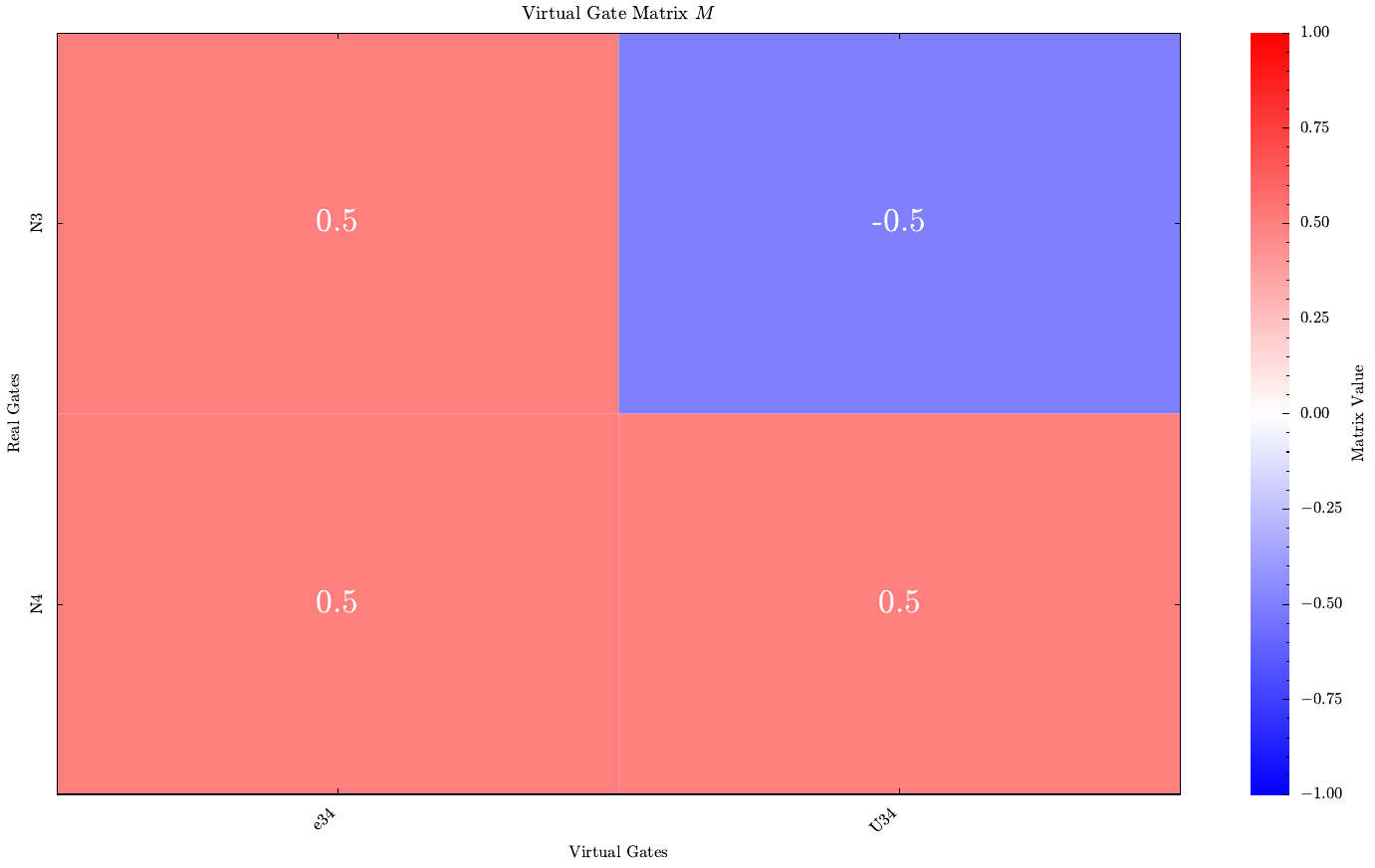}
    \caption{Detuning $\epsilon_{34}$ and Potential energy $U_{34}$ virtual gate matrix.}
    \label{}
\end{figure}

\begin{figure}[H]
    \centering
    \includegraphics[width =0.8\linewidth]{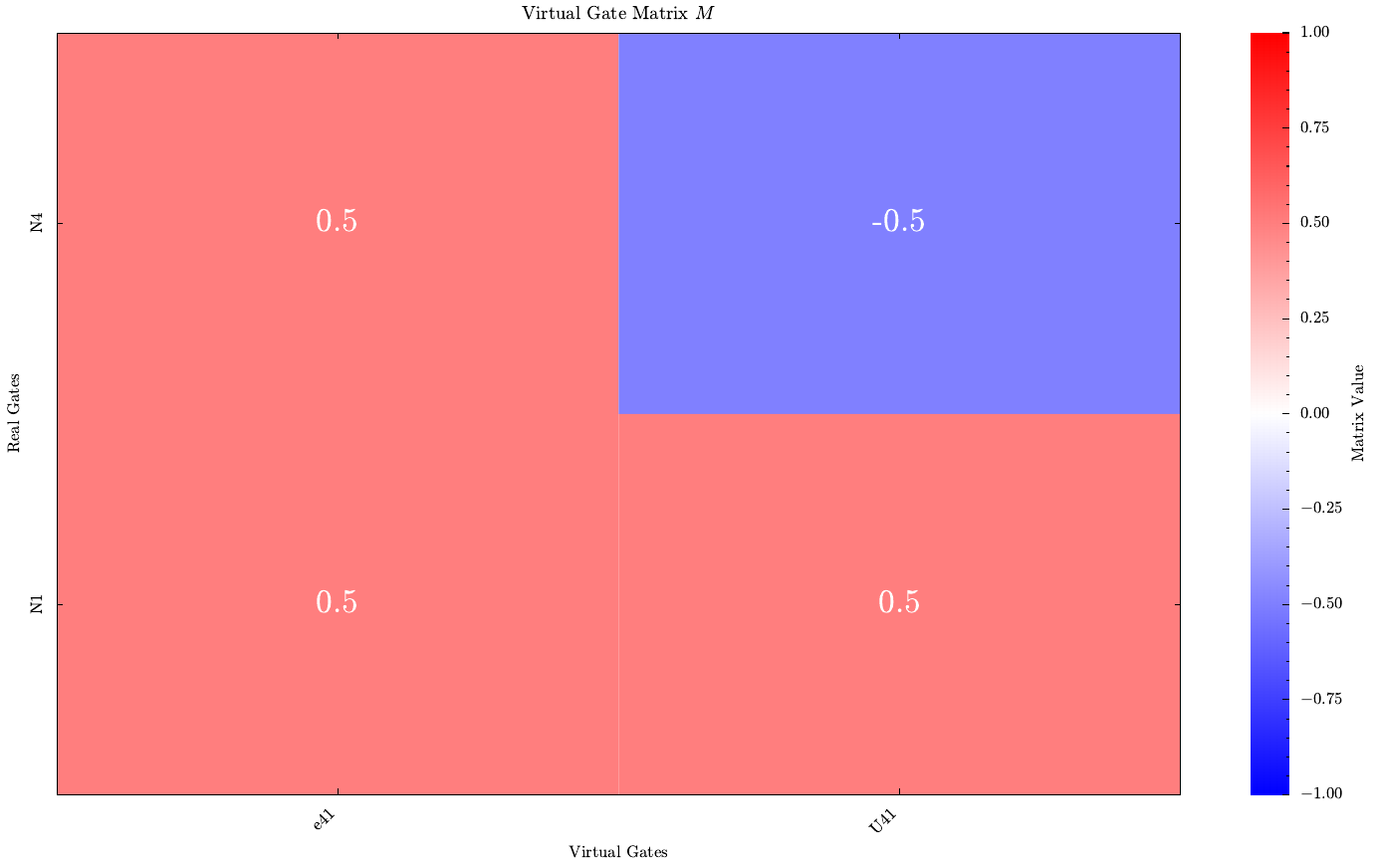}
    \caption{Detuning $\epsilon_{41}$ and Potential energy $U_{41}$ virtual gate matrix.}
    \label{}
\end{figure}

\section{Manual fitting of Charge Stability Diagrams for Triangulation}
\label{supp_manualfitting}

\subsection{Triangulation}
\label{triangulation_method}

In order to find the relative lever arms of the gates to the quantum dots, we collect charge stability diagrams of the plunger gate \textit{P\textsubscript{i}} vs all surrounding gates \textit{G\textsubscript{j}}. The full set of diagrams, fitted and unfitted are shown below. The diagrams are centered around a particular reservoir transition line for the gate being swept. The plots are then manually fitted, ensuring that all of the transitions correspond to the same intersection along the y=0 axis.

We collect data of all of the gates vs the virtual gate $L_{I}$, where i= 1,2,3,4. The L gates are virtualised such that all the surrounding gates are virtualised to the lower quantum dot charge transition. We then manually fit the charge transition lines on all the charge stability diagrams, by categorizing the slopes into 'steep' (red), 'less steep' (blue), and 'interdot transition' lines (purple). When the gate is on the $y = 0$, we correlate the charge transitions across the different fits and ensure they are self-consistent. We then separately find the averages of the slopes $\bar{\alpha_G}$. As we do not measure the absolute lever arm, we must find the lever arm relative to a common gate. We choose $L_i$, which is swept on the $x$-axis, as this common gate and find the relative lever arm to be
\begin{equation}
    \alpha = \frac{\bar{\alpha_G}}{\bar{\alpha_{L_{i}}}}
\end{equation}\label{eqn:relativeleverarm}

We normalize to the plunger gate which is swept on the $x$-axis, and show the spread in the slopes.

\begin{figure}
    \centering
    \includegraphics[width=\linewidth]{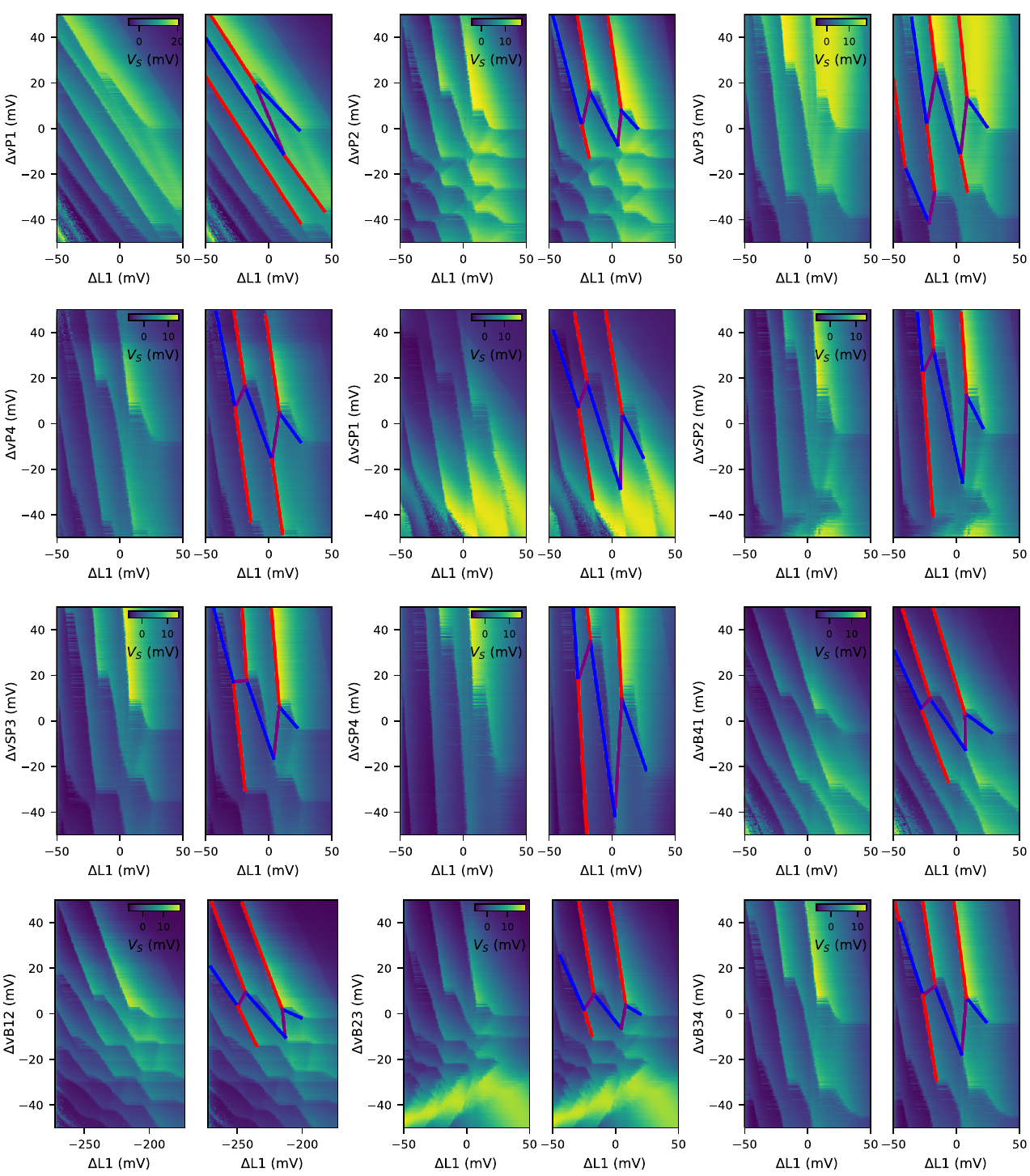}
    \caption{Triangulation data and the fitted triangulation data for the double dot pair under plunger gates P\textsubscript{1} and P\textsubscript{2}}
    \label{fig:L1L2fitted unfitted}
\end{figure}

\begin{figure}[H]
    \centering
    \includegraphics[width=\linewidth]{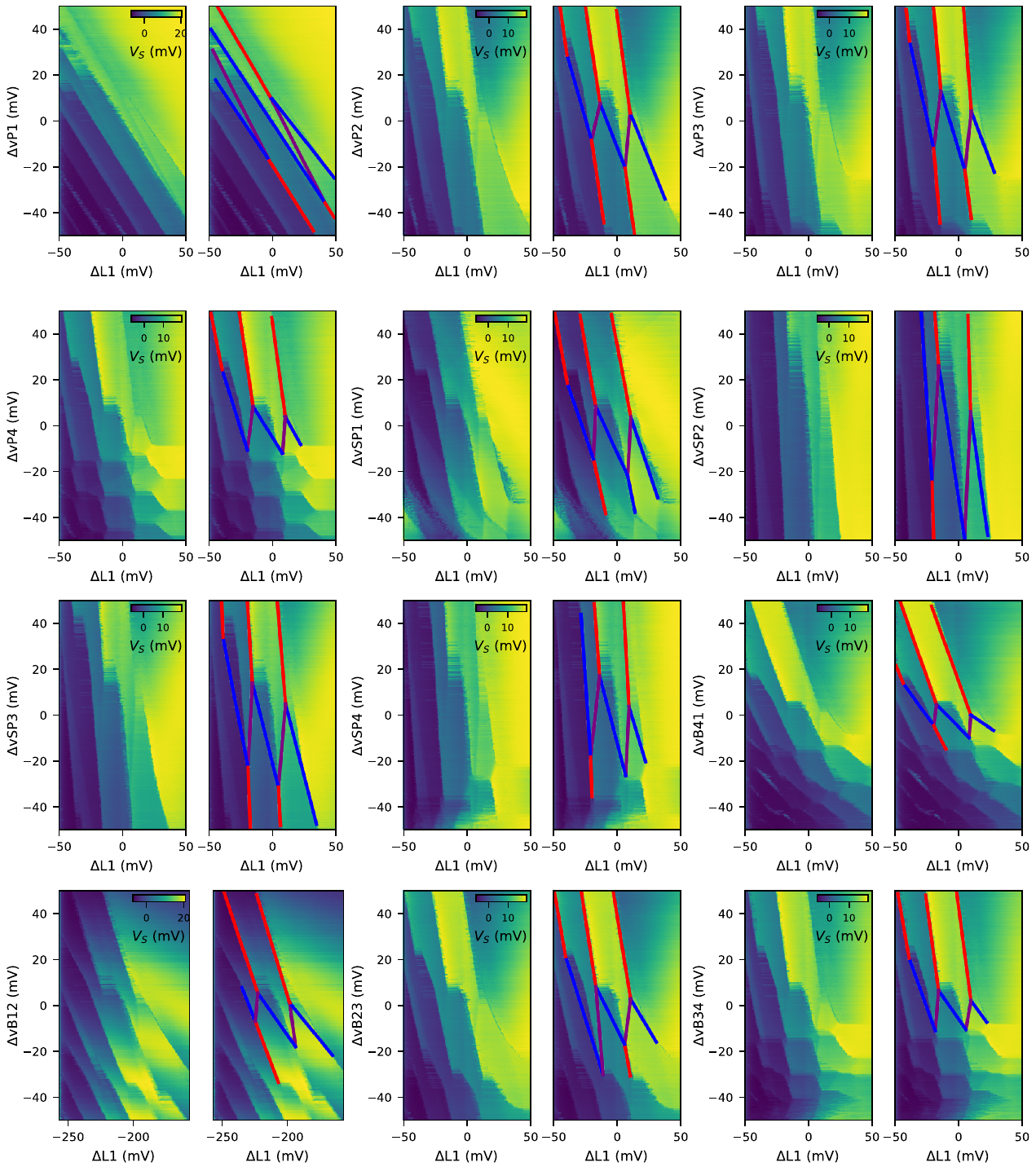}
    \caption{Triangulation data and the fitted triangulation data for the double dot pair under plunger gates P\textsubscript{1} and P\textsubscript{4}}
    \label{fig:L1L4fitted unfitted}
\end{figure}

\begin{figure}[H]
    \centering
    \includegraphics[width=\linewidth]{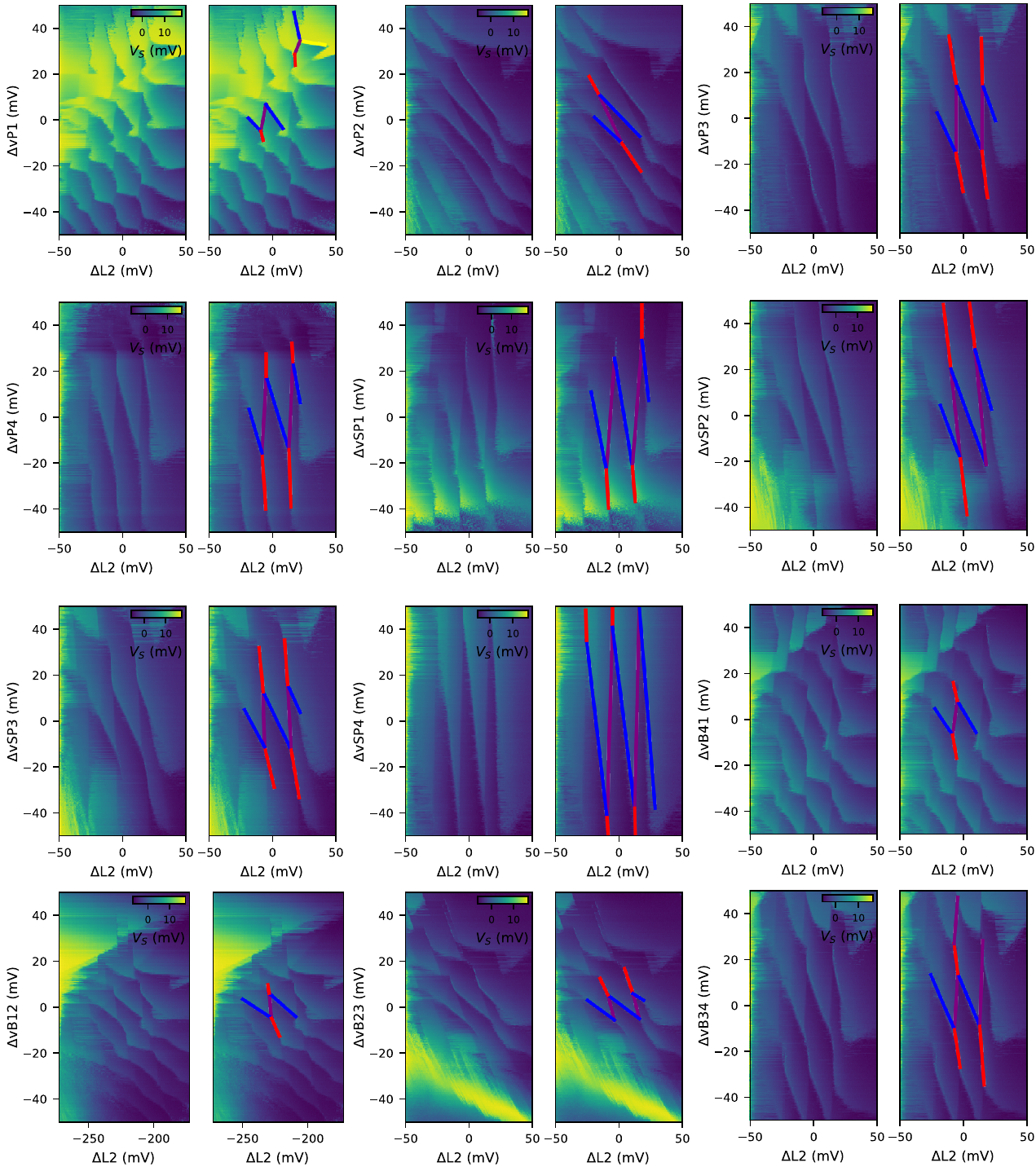}
    \caption{Triangulation data and the fitted triangulation data for the double dot pair under plunger gates P\textsubscript{2} and P\textsubscript{1}}
    \label{fig:L2L1fitted unfitted}
\end{figure}

\begin{figure}[H]
    \centering
    \includegraphics[width=\linewidth]{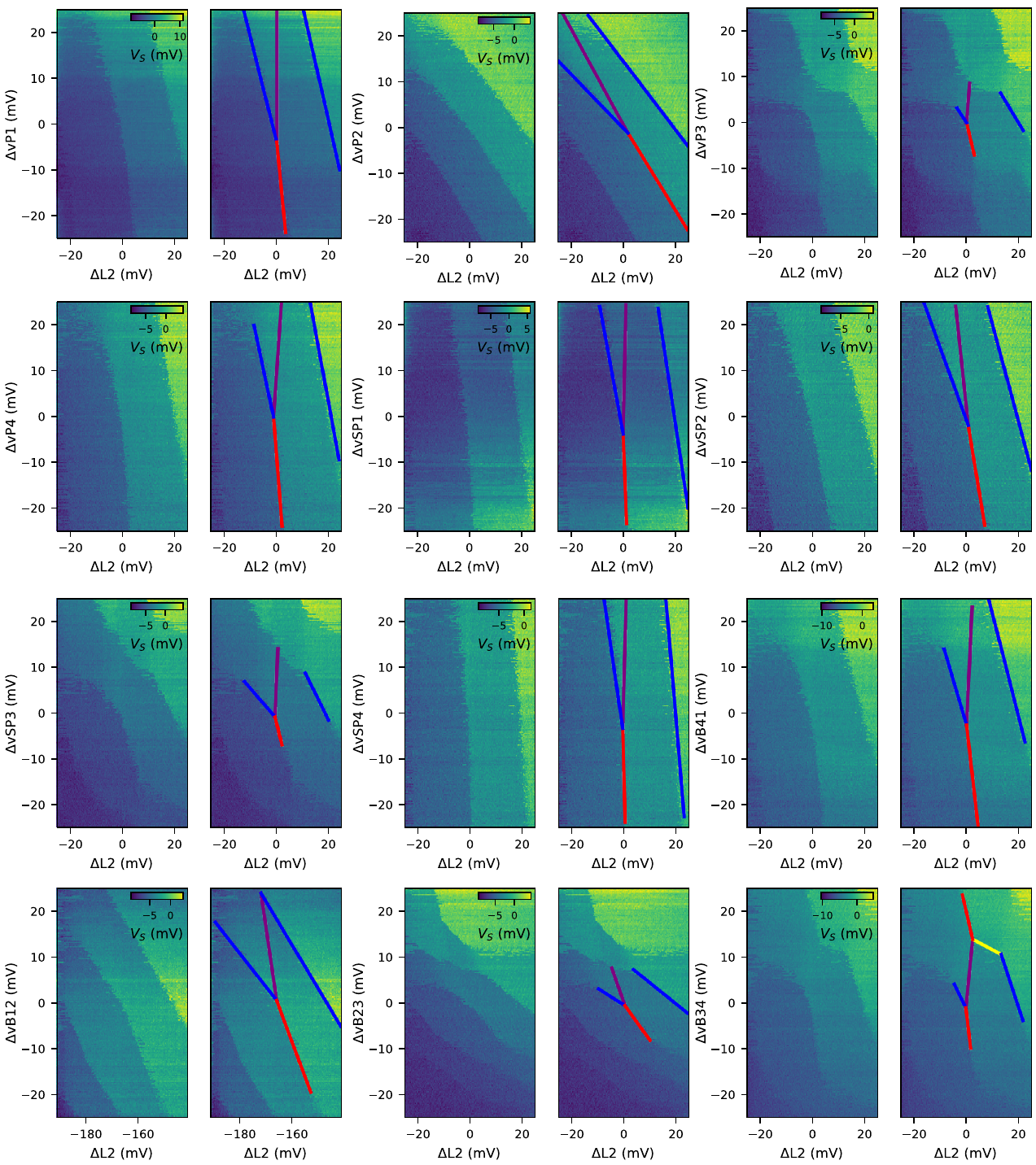}
    \caption{Triangulation data and the fitted triangulation data for the double dot pair under plunger gates P\textsubscript{2} and P\textsubscript{3}}
    \label{fig:L2L3fitted unfitted}
\end{figure}

\begin{figure}[H]
    \centering
    \includegraphics[width=\linewidth]{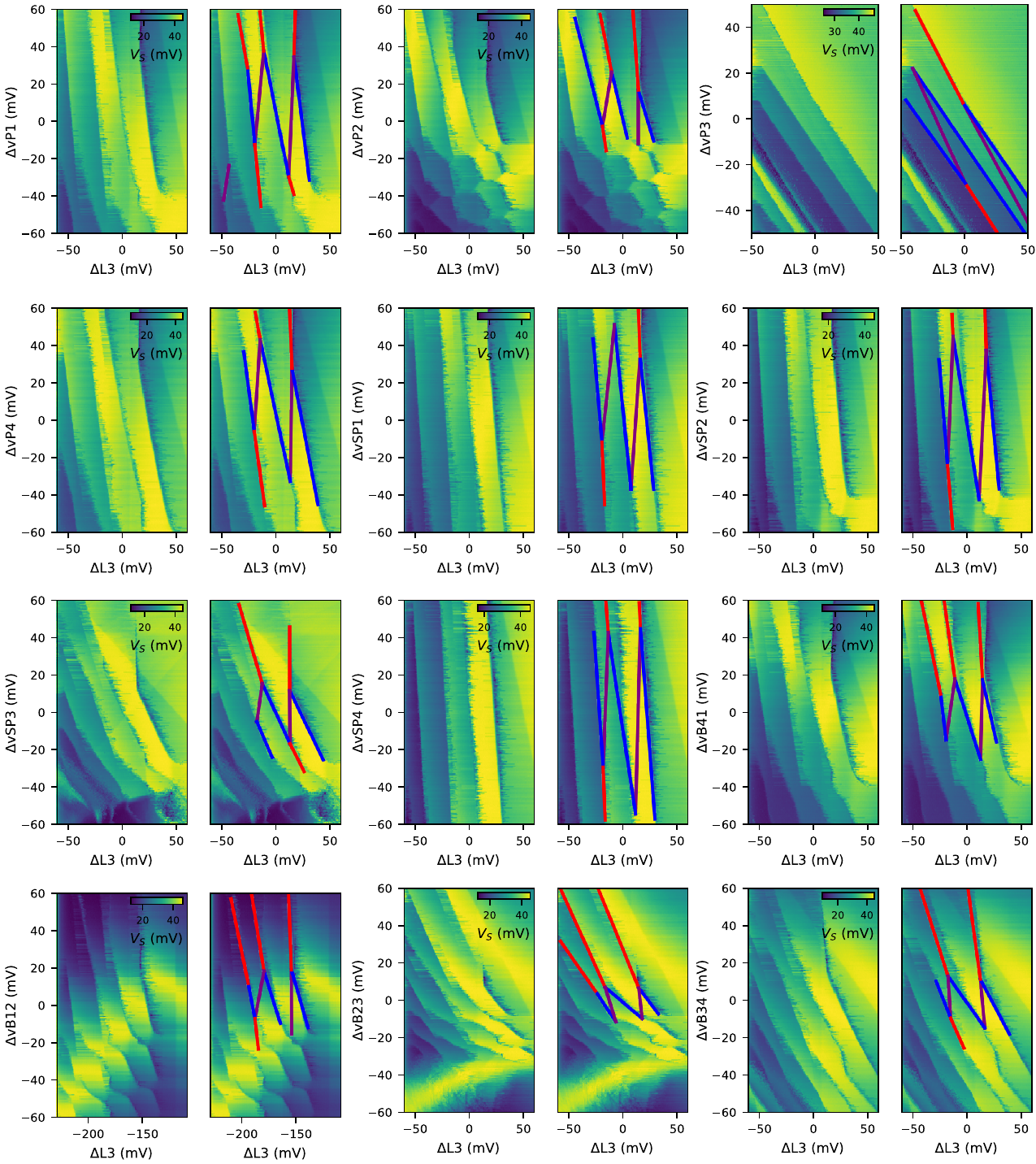}
    \caption{Triangulation data and the fitted triangulation data for the double dot pair under plunger gates P\textsubscript{3} and P\textsubscript{2}}
    \label{fig:L3L2fitted unfitted}
\end{figure}

\begin{figure}[H]
    \centering
    \includegraphics[width=\linewidth]{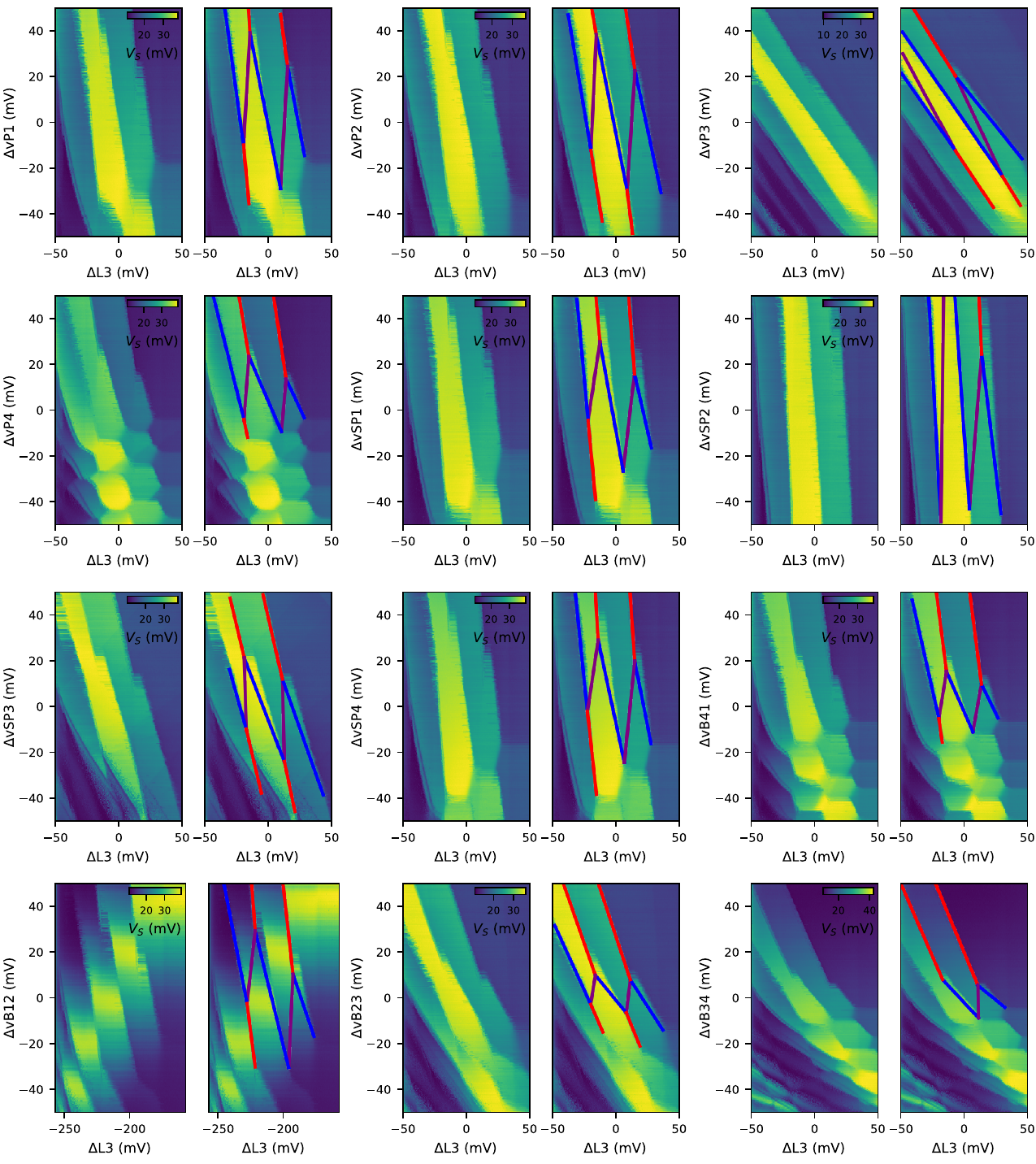}
    \caption{Triangulation data and the fitted triangulation data for the double dot pair under plunger gates P\textsubscript{3} and P\textsubscript{4}}
    \label{fig:L3L4fitted unfitted}
\end{figure}

\begin{figure}[H]
    \centering
    \includegraphics[width=\linewidth]{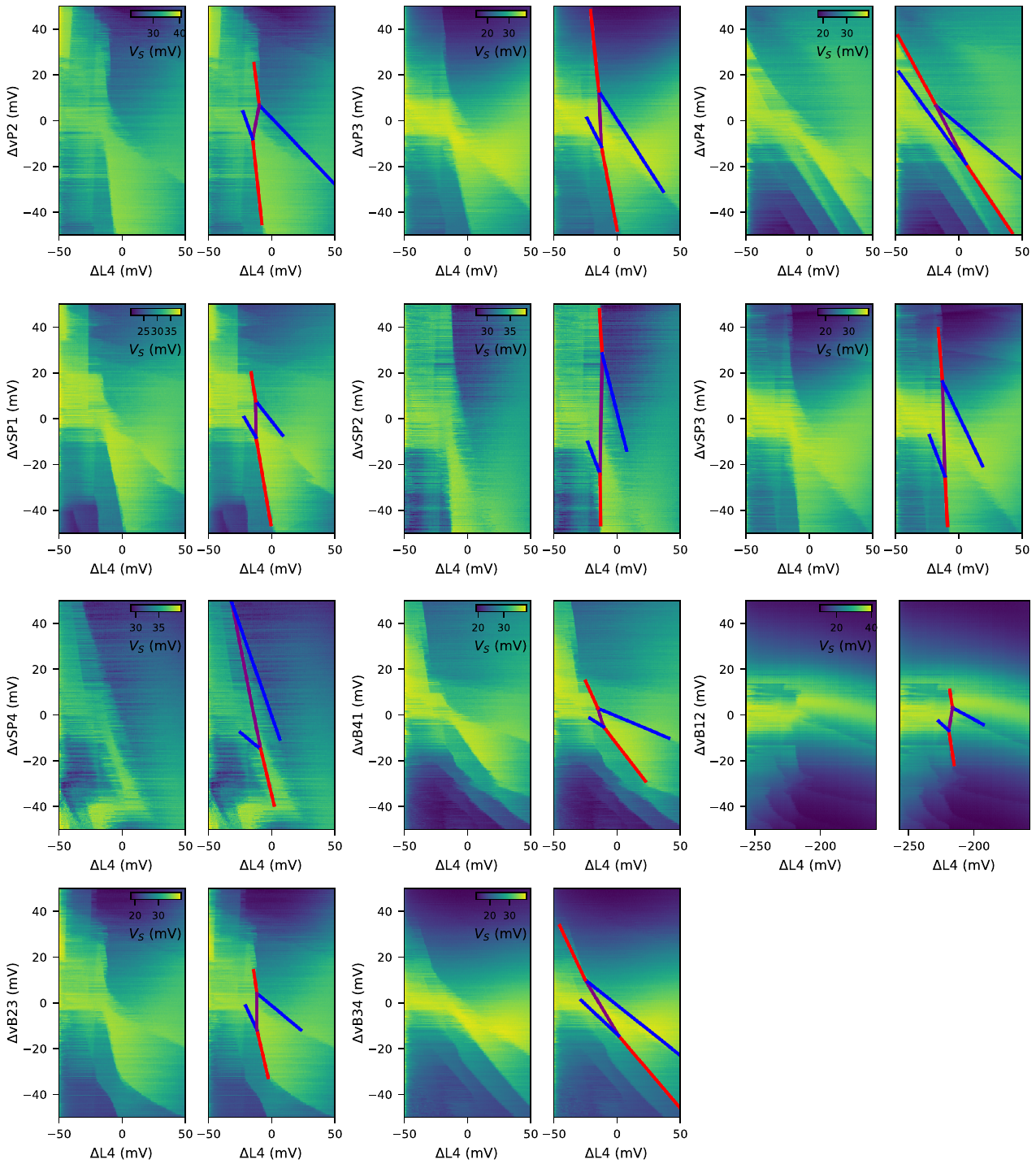}
    \caption{Triangulation data and the fitted triangulation data for the double dot pair under plunger gates P\textsubscript{4} and P\textsubscript{1}}
    \label{fig:L4L1fitted unfitted}
\end{figure}

\begin{figure}[H]
    \centering
    \includegraphics[width=\linewidth]{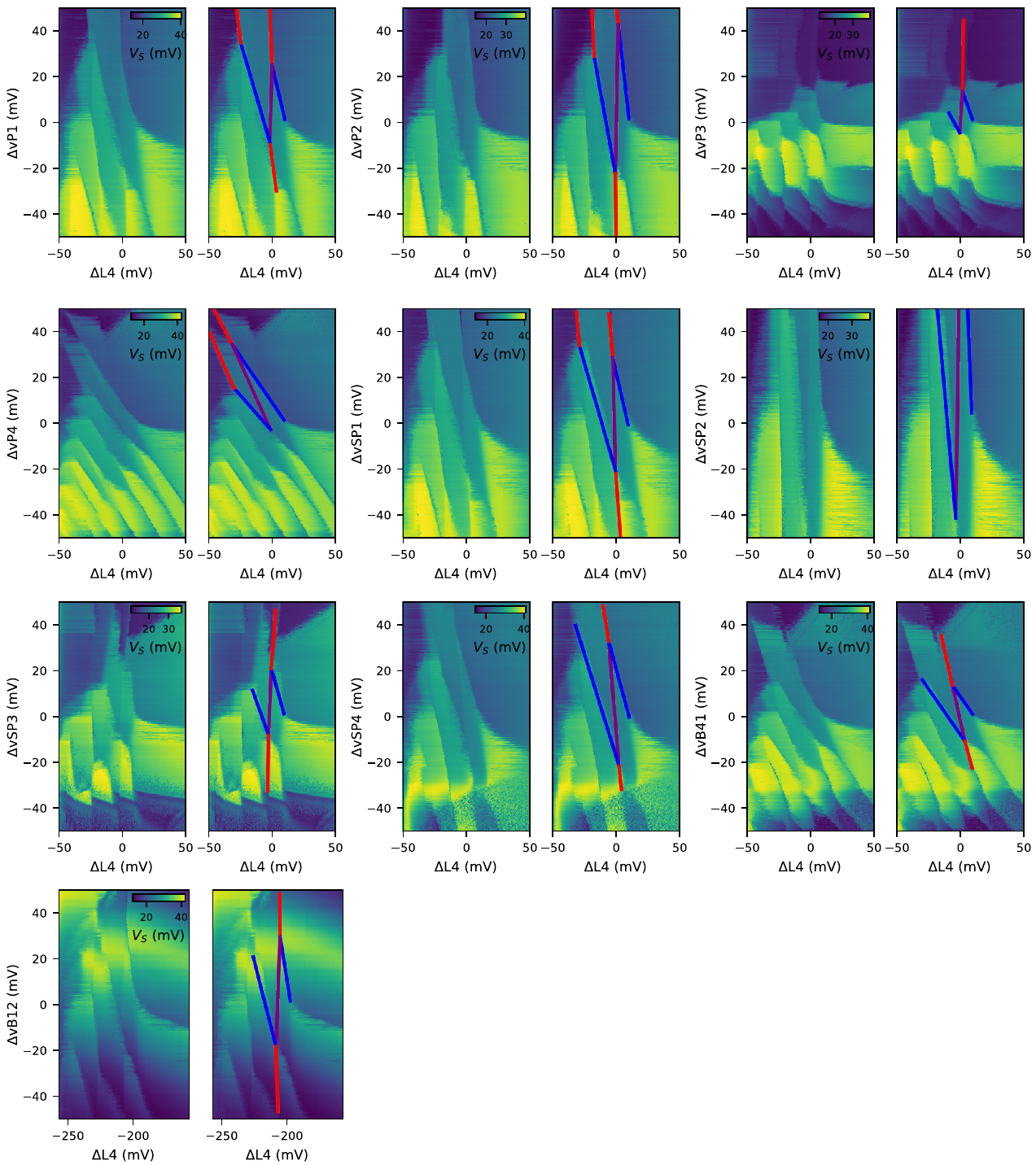}
    \caption{Triangulation data and the fitted triangulation data for the double dot pair under plunger gates P\textsubscript{4} and P\textsubscript{3}}
    \label{fig:L4L3fitted unfitted}
\end{figure}

\begin{figure}[H]
    \centering
    \includegraphics[width=\linewidth]{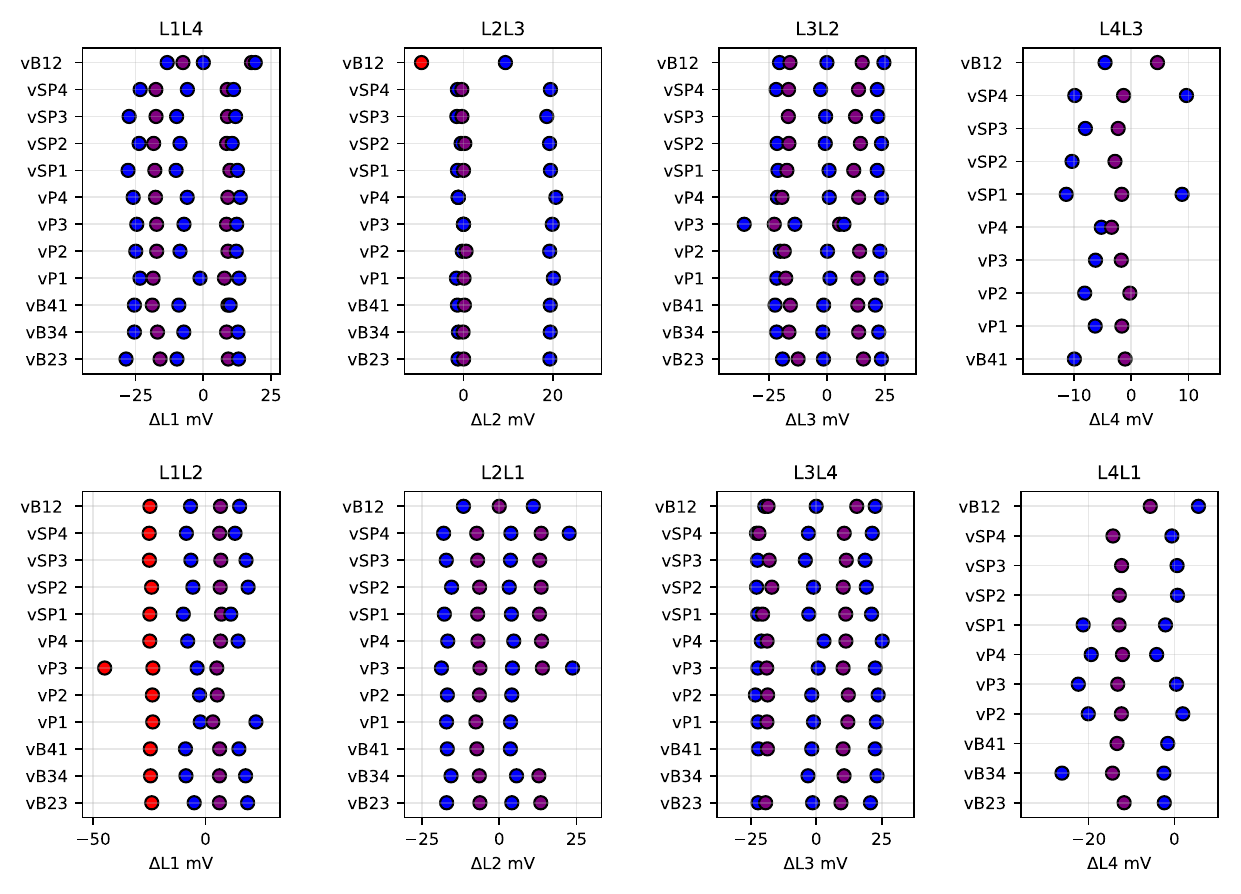}
    \caption{The intersection points along y=0 for all the fitted data sets.}
    \label{Supp:intersectionpts}
\end{figure}

In Fig.\ref{Supp:intersectionpts} we show the intersection points for each charge stability diagram along the linecut of $y=0$. We do this to ensure that the line categorisation is consistent across charge stability diagrams. 

In the figures below, we plot the spread of the relative lever arms of the gates. In some, we see that nearby barrier gates have a larger coupling to the bottom quantum dot than the plunger gate. This is consistent with schroedinger-poisson simulation results reported in \cite{Tidjani2023}, where the large spread in the wavefunction of the lower quantum dot due to decreased confinement, and some screening from the charges in the upper quantum dot, reduces the coupling of the plunger gate to the lower quantum dot. Additionally, the barrier gates have a non-linear effect on the lower quantum dot in particular, moving it as their voltages change. This means there can be large differences in the slopes between transitions. 


Additionally, there is further complication for the gates which do not have direct reservoir transitions being P2 and P4. As they load via the dots closest to the sensor (P1 and P3), when swept on the fast axis they have strong latching features. Therefore, we centre the diagram on a known transition, then flip the axis. In this way, we know which transition we are at on our DC point, and then can categorise the other slopes based on their steepness. This also decreases the number of transitions we can reliably fit.

\begin{figure}[H]
    \centering
    \includegraphics[width=\linewidth]{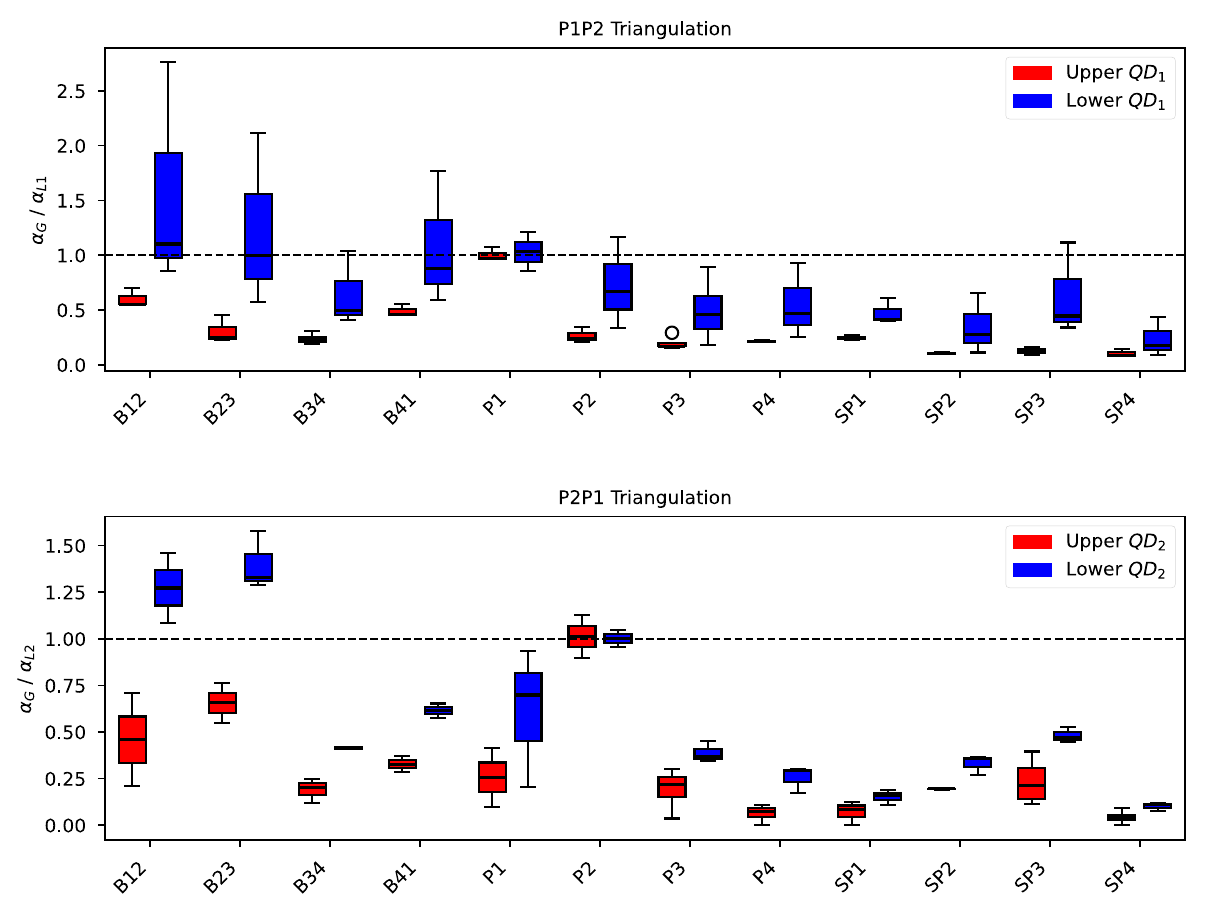}
    \caption{Boxplots indicating the spread in the relative lever arm across the various transitions. }
    \label{fig:SUPPtri_CSDs_fit_p1p2}
\end{figure}

\begin{figure}[H]
    \centering
    \includegraphics[width=\linewidth]{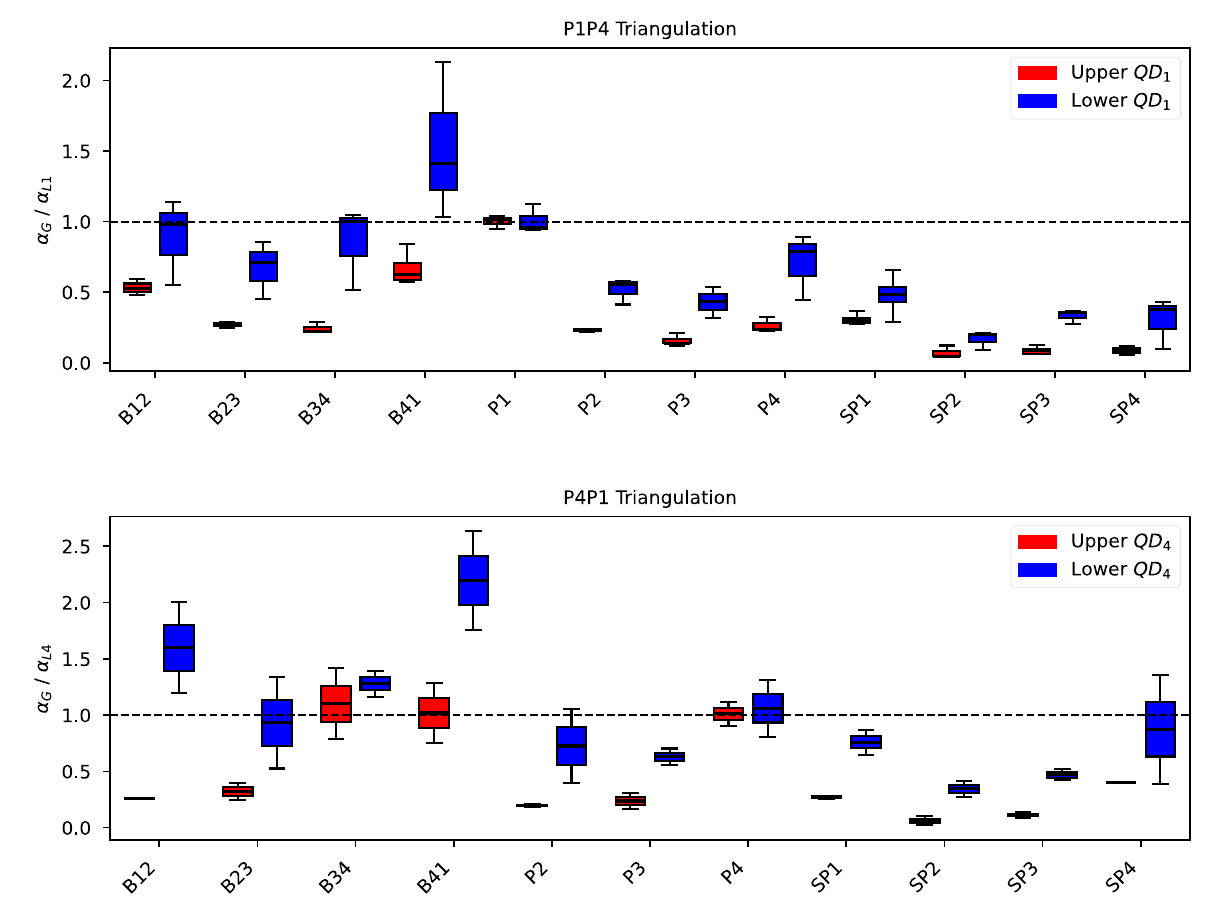}
    \caption{Boxplots indicating the spread in the relative lever arm across the various transitions. }
    \label{fig:SUPPtri_CSDs_fit_p1p4}
\end{figure}

\begin{figure}[H]
    \centering
    \includegraphics[width=\linewidth]{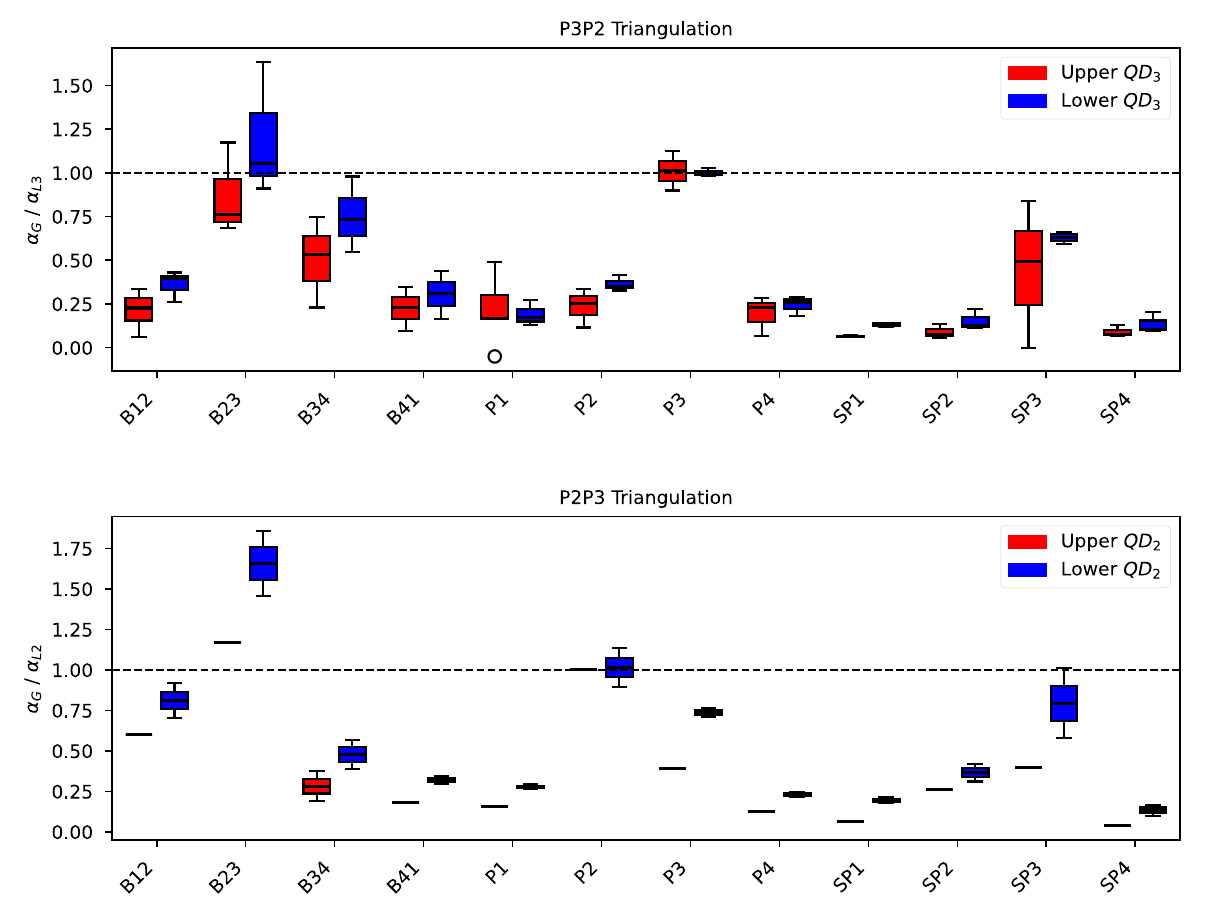}
    \caption{Boxplots indicating the spread in the relative lever arm across the various transitions. }
    \label{fig:SUPPtri_CSDs_fit_p3p2}
\end{figure}

\begin{figure}[H]
    \centering
    \includegraphics[width=\linewidth]{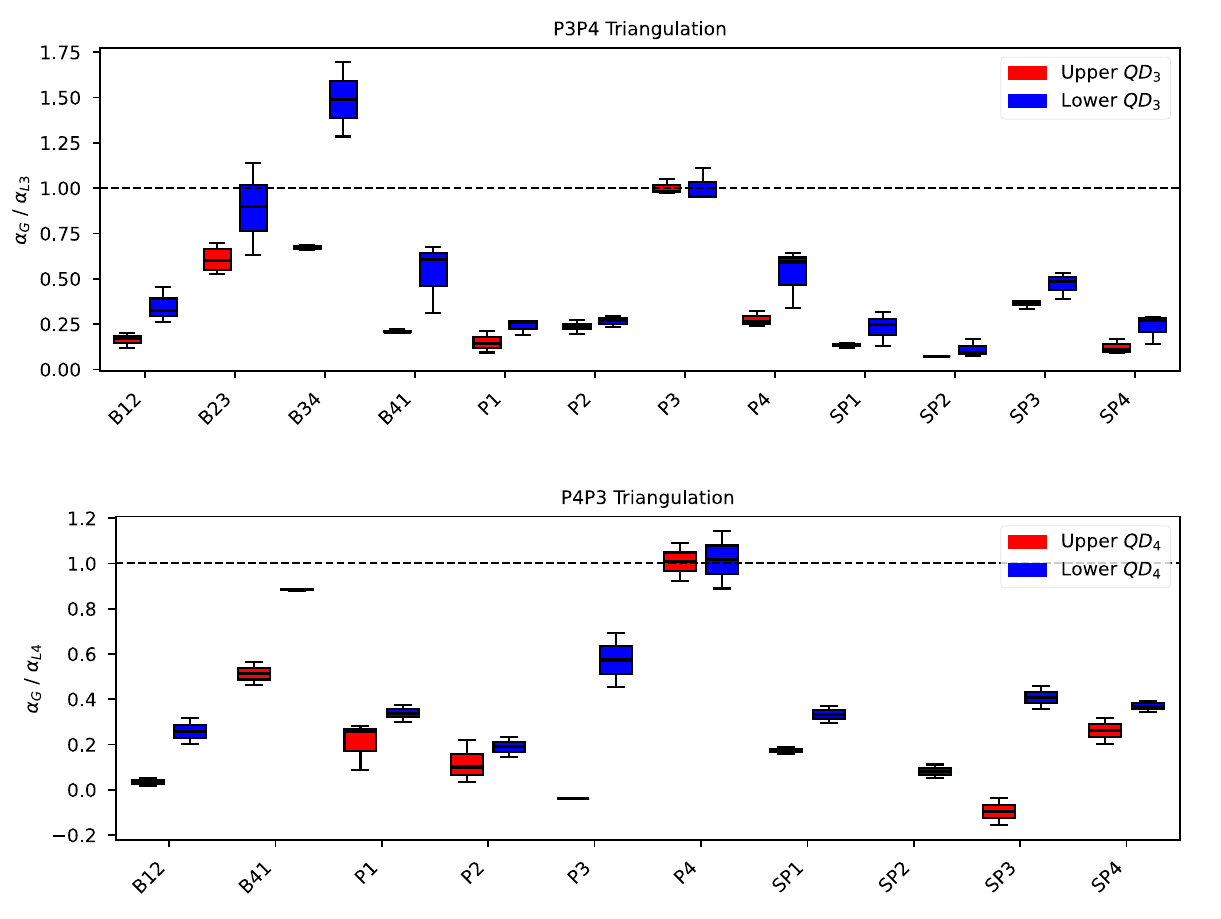}
    \caption{Boxplots indicating the spread in the relative lever arm across the various transitions. }
    \label{fig:SUPPtri_CSDs_fit_p3P4}
\end{figure}

\section{Capacitance modelling}\label{qarray_sim}
The charge stability diagrams observed in experiments and presented can be reproduced using a capacitance simulation. 
We employ Qarray, an open source package\cite{qarray}, to simulate charge stability diagrams through the constant capacitance model.
Normally, for a simple simulation such as of a double quantum dot, we require two plunger gates under which the two quantum dots are formed. 
The number of plunger gates is equal to the number of quantum dots.
However, in a device with a bilayer heterostructure, we can now have that two quantum dots couple primarily to the same plunger, i.e. a vertical double quantum dot, while no quantum dot(s) are formed underneath the other plunger gate(s).
Nevertheless, for the simulations we will still consider all four plunger gates.
Firstly, because of the virtualization used in experiments, which we include in the simulation as well.
Second, because we also use the unaccumulated plunger gates in simulation as levers, as alluded to in the main manuscript, to detune the chemical potential of the lower layer with respect to the upper.
That is, to move the loading lines of the lower dots relative to the loading lines of the upper dots in the simulated charge stability diagram.

The Qarray simulations require capacitance matrices C\textsubscript{dd} and C\textsubscript{gd} as input, which we determine by manually fitting the charge stability diagram. 
Here, we show these capacitance matrices for the fit in Fig.\ref{fig2:simulate_exp_4dot}, in which two quantum dots are formed in the upper and lower quantum well, underneath plunger gate 2 and 3 each.
The capacitive couplings between the four dots are contained within the capacitance matrix C\textsubscript{dd}: 
 \begin{equation}
    C_{dd}=
\begin{bNiceMatrix}[last-col]
0& 0.00267& 0.016& 0.00133& \text{2\textit{u}}\\
0.00267& 0& 0.00133& 0.016& \text{3\textit{u}}\\
0.016& 0.00133& 0& 0.008& \text{2$\ell$}\\
0.00133& 0.016& 0.008& 0& \text{3$\ell$}\\
\end{bNiceMatrix}
\label{Eq: Cdd matrix}
\end{equation}
The inputs to this matrix mainly determine the length of the interdot transitions between different quantum dots. 

C\textsubscript{gd} is used to represent the capacitive couplings of the four plungers to the four quantum dots underneath P\textsubscript{2} and P\textsubscript{3}:
 \begin{equation}
    C_{gd}=
\begin{bNiceMatrix}[first-row, last-col]
P\textsubscript{1} & P\textsubscript{2} & P\textsubscript{3} & P\textsubscript{4} \\
0.00267& 0.03466& 0.00267& 0.0& \text{2\textit{u}}\\
0.0& 0.00267& 0.03466& 0.00267& \text{3\textit{u}}\\
0.00533& 0.01067& 0.00533& 0.0& \text{2$\ell$}\\
0.0& 0.00533& 0.01067& 0.00533& \text{3$\ell$}\\
\end{bNiceMatrix}
\label{Eq: Cgd matrix}
\end{equation}
It primarily determines the slope and spacing between transition lines in the charge stability diagram. 

The simulations using the manually fitted capacitance matrices are qualitatively in agreement with the experimentally measured charge stability diagrams.
Although these capacitance matrices differ for each simulation, we observe some trends between them.
One is that the largest scale in $C_{dd}$ seems to be capacitive coupling between quantum dots on top of each other followed by coupling between quantum dots in the lower layer. 
First, this indicates that quantum dots in different quantum wells are formed at a similar positions, below the plunger gates.
Moreover, it also means that quantum dots in the lower quantum well feel each other more strongly than if they were in the upper quantum well.
This conforms to our expectation, as we have seen from previous work that the wavefunctions of holes in the lower quantum dots are less localised\cite{Tidjani2023}. 
Furthermore, we note from the fitted C\textsubscript{gd} that the coupling of the plunger to the upper quantum dot is larger than the lower one, which is again in agreement with previous work \cite{Tidjani2023}.
The argument for this observation is that in the perspective of the lower quantum dot, the upper quantum dot partially screens the electric field generated by the plunger gate.
This leads to a reduced lever arm of the lower quantum dot.
This implies again that the wavefunction is more spread out in contrast to the upper quantum dots which are more localized under their respective plunger gate. 
Nevertheless, the coupling of the lower dot seems to be largest under its respective plunger gate indicating that this lower dot is formed underneath the plunger and not in between plungers gates. 
We note that because of the blurring/broadening of transition lines and latching, there is a (small) error margin in the fitted capacitance matrices.
However, the error in the capacitances does not significantly change its value, or rather, more importantly, does not change the order between the different capacitances.

Differences between measured and simulated charge stability diagrams can also be attributed to the inherent limitations of the constant capacitance model. 
This includes the assumption that the capacitances do not vary as a function of gate voltage, while we do see in e.g. Fig.\ref{fig2:simulate_exp_4dot} that the charging energy decreases at higher occupation.
In addition, it does not capture quantum mechanical effects such as tunneling which is known to bend transition lines \cite{Hanson2007SpinsDots}.
This can also lead to differences between the experimentally measured and simulated charge stability diagram, especially if the lower quantum dots are strongly tunnel coupled to each other.




\end{appendices}

\normalem

\end{document}